\def\draftversion{false}

\RequirePackage{ifthen}
\ifthenelse{\equal{\draftversion}{true}}{
  \documentclass[prb,galley,showpacs,preprintnumbers,citeautoscript,
      amsmath,amssymb]{revtex4}
}{
  \documentclass[citeautoscript,floatfix,aps,prb,twocolumn,
      superscriptaddress]{revtex4-1}
}

\usepackage{amsmath}
\usepackage{graphicx}
\usepackage{epstopdf}
\usepackage{natbib}
\usepackage{array}
\usepackage{dcolumn}
\usepackage{bm}

\usepackage{soul}  

\usepackage[usenames,dvipsnames]{color}

\ifthenelse{\equal{\draftversion}{true}}{
  \marginparwidth 2.7in
  \marginparsep 0.5in
  \newcounter{comm} 
  \def\commnext{\stepcounter{comm}}
  \def\commtext{{\bf\color{blue}[\arabic{comm}]}}
  \def\commmar{{\bf\color{blue}[\arabic{comm}]}}
  \def\dvm#1{\commnext\marginpar{\small DV\commmar: #1}\commtext}
  \def\cdm#1{\commnext\marginpar{\small CED\commmar: #1}\commtext}
  \def\msm#1{\commnext\marginpar{\small MS\commmar: #1}\commtext}
  \def\mlab#1{\marginpar{\small\bf #1}}
  \def\tnewpage{\newpage\marginpar{\small Temporary newpage}}
}{
  \def\dvm#1{}
  \def\cdm#1{}
  \def\msm#1{}
  \def\mlab#1{}
  \def\tnewpage{}
}

\begin{document}
\title{Current-density implementation for calculating flexoelectric coefficients}
\author{Cyrus E. Dreyer}
\affiliation{Department of Physics and Astronomy, Rutgers University, Piscataway, New Jersey 08845-0849, USA}
\author{Massimiliano Stengel}
\affiliation{ICREA-Instituci\'{o} Catalana de Recerca i Estudis Avan\c{c}ats, 08010 Barcelona, Spain}
\affiliation{Institut de Ci\`{e}ncia de Materials de Barcelona (ICMAB-CSIC), Campus UAB, 08193 Bellaterra, Spain}
\author{David Vanderbilt}
\affiliation{Department of Physics and Astronomy, Rutgers University, Piscataway, New Jersey 08845-0849, USA}
\date{\today}

\begin{abstract}
  The flexoelectric effect refers to polarization induced in an
  insulator when a strain gradient is applied. We have developed a
  first-principles methodology based on density-functional
  perturbation theory to calculate the elements of the bulk
  clamped-ion flexoelectric tensor. In order to determine the
  transverse and shear components directly from a unit cell
  calculation, we calculate the current density induced by the
  adiabatic atomic displacements of a long-wavelength acoustic
  phonon. Previous implementations based on the charge-density
  response required supercells to capture these components.  At the
  heart of our approach is the development of an expression for the
  current-density response to a generic long-wavelength phonon
  perturbation that is valid for the case of nonlocal
  pseudopotentials.  We benchmark our methodology on simple systems of
  isolated noble gas atoms, and apply it to calculate the clamped-ion
  flexoelectric constants for a variety of technologically important
  cubic oxides.
\end{abstract}

\maketitle

\section{Introduction}

The flexoelectric (FxE) effect, where polarization is induced by a
strain gradient, is universal in all insulators. As devices shrink to
the micro and nano scale, large strain gradients can occur, and
therefore the FxE effect can play a significant role in the properties
of such devices, influencing the so-called dielectric dead
layer\cite{Majdoub2009}, domain walls and domain
structure\cite{Lee2011,Yudin2012,Borisevich2012}, relative
permittivity and Curie temperature\cite{Catalan2004,Catalan2005},
critical thickness of films to exhibit switchable
polarization\cite{Zhou2012}, and spontaneous polarization in the
vicinity of twin and antiphase boundaries\cite{Morozovska2012}.  Also,
the FxE effect can be exploited for novel device design paradigms,
such as piezoelectric ``meta-materials'' constructed from
nonpiezoelectric constituents\cite{Zhu2006,Bhaskar2016}, or mechanical
switching of ferroelectric polarization \cite{Lu2012,Gruverman2003}.

One of the crucial limitations to understanding and exploiting the FxE
effect is the lack of a clear experimental and theoretical consensus
on the size and sign of the FxE coefficients, even in commonly studied
materials such at SrTiO$_3$ and BaTiO$_3$\cite{Zubko2013,Yudin2013}. A
key element to forming this understanding is the development of an
efficient first-principles methodology to calculate all of the
components of the bulk FxE tensor. Recently, Stengel,
\cite{Stengel2013} and Hong and Vanderbilt\cite{Hong2011,Hong2013}
(HV), developed the formalism for calculating the full bulk FxE tensor
from first principles.
\footnote{The FxE response of any finite crystal also has an important
  contribution from the surface, as discussed in
  Refs.~\onlinecite{Stengel2013natcom} and
  \onlinecite{StengelChapter}, and calculated using density-functional
  theory for SrTiO$_3$ in \onlinecite{Stengel2014}. In this work, we
  will exclusively focus on bulk contribution, which poses a more
  significant challenge for a computational treatment.}

Each element of the FxE tensor has a ``clamped-ion'' (CI) contribution,
arising from the effect of the strain gradient on the valence
electrons in the crystal, and a ``lattice-mediated'' (LM) contribution,
arising from internal relaxations induced by the applied strain and strain
gradient \cite{Stengel2013,Hong2013}. 
In Refs.~\onlinecite{Hong2011} and \onlinecite{Hong2013}, HV described
an implementation for calculating the bulk CI and LM longitudinal FxE
coefficients (i.e., the coefficients relating the induced polarization
in direction $\alpha$ to a gradient of uniaxial strain
$\varepsilon_{\alpha\alpha}$, also in direction $\alpha$). Their
methodology involved using density functional theory (DFT) to
calculate the real-space response of the charge density to atomic
displacements in a simple $N\times 1\times 1$ bulk supercell containing
$N$ repitions of the primitive bulk cell.

In Ref.~\onlinecite{Stengel2014}, Stengel developed a strategy
that allowed a calculation of the full FxE response for cubic SrTiO$_3$
based in part on the charge-density response to a long-wavelength
acoustic phonon, and in part on large slab supercell calculations
(repeated slabs separated by vacuum). The first part of this methodology
allowed the LM contributions to all bulk FxE tensor elements,
as well as the CI contributions to the longitudinal coefficients,
to be determined from linear-response calculation on a single unit
cell using density-functional perturbation theory (DFPT)
\cite{Baroni2001}. However, the ``transverse'' and ``shear'' CI
contributions \cite{Hong2013,Stengel2014,Stengel2013natcom} had to
be calculated indirectly by relating them to the open-circuit
electric field appearing across the slab when a long wavelength
acoustic phonon was applied to the slab supercell as a whole.
As a result, this implementation required DFPT calculations to be
performed on large slab supercells.

The implementation described in Ref.~\onlinecite{Stengel2014} thus
  provides a methodology for calculating the full FxE
  tensor for a given material.  However, the reliance on
  computationally intensive slab supercell calculations for the
  transverse and shear CI coefficients represents a significant
  limitation to efficient calculation, especially in complex
  materials. Therefore, it is highly desirable to develop an
  approach that allows the full bulk FxE tensor, including its
  longitudinal, transverse, and shear components, to be obtained
  from DFPT calculations on single unit cells.

The essential problem is that single-unit-cell DFPT calculations
that determine only the charge-density response to a long-wavelength
phonon, as in Ref.~\onlinecite{Stengel2014}, are incapable of
revealing the transverse and shear CI contributions, since the
induced charge is proportional to the \textit{divergence} of the
polarization, which is absent for transverse phonons.  To go
further, it is necessary to compute the induced \textit{polarization
itself}.  Unfortunately, the well-known Berry-phase formulation
\cite{KingSmith1993,Resta1994} of the electric polarization is
useless here, since it provides only the total polarization, which
averages to zero over a phonon wavelength. Instead, we need access
to the spatially resolved polarization on the scale of the
wavelength.  The only clear path to obtaining this local
polarization is via its relation to the adiabatic current
density \cite{Hong2013,Stengel2013,StengelUNPUB}.  Thus, the desired
methodology is one that computes the spatially resolved
\textit{ current density} induced by a strain gradient perturbation
\cite{Hong2013,Stengel2013,StengelUNPUB} in the context of
long-wavelength longitudinal \textit{and transverse} phonons.

The microscopic current density is, of course, just
  proportional to the quantum-mechanical probability current, as
  discussed in any standard textbook \cite{Sakuri}.  However, this
  standard formula assumes a local Hamiltonian of the form
  $H=p^2/2m+V$ with a local potential $V$. Thus, it becomes problematic if
  the Hamiltonian of interest contains \emph{nonlocal} potentials,
  as the probability current no longer satisfies the continuity
  equation\cite{Li2008}.  This issue is very relevant in the context of
  DFT, since most popular implementations make use of a plane-wave
  basis set with a pseudopotential approximation to reduce the size
  of the basis set by avoiding an explicit description of the core
  electrons.  Virtually all modern pseudopotential implementations
  contain nonlocal potentials in the form of projectors that operate
  on the wavefunctions
  \cite{Vanderbilt1990,Hamann1979,Kleinman1982,Blochl1994}.
  Therefore, the standard formula for the current density is not a
  fit starting point for the current-response theory that we have in
  mind (we expand on these considerations in Sec.~\ref{curden}).

The definition and calculation of the microscopic current density in a
nonlocal pseudopotential context is a rather general problem that has
received considerable previous attention
\cite{Umari2001,Li2008,Vignale1991,ICL2001,Pickard2003,Mauri1996,Mauri1996_nmr}
in view of its application to the calculation of magnetic
susceptibility
\cite{Vignale1991,ICL2001,Pickard2003,Mauri1996,Mauri1996_nmr},
nuclear magnetic resonance chemical shifts \cite{Pickard2001},
electron paramagnetic resonance $g$ tensors\cite{Pickard2002}, and so
forth.  Unfortunately a general, systematic solution that is
appropriate to our scopes has not emerged yet.  To see why this is
challenging, it is important to note that the continuity equation is
only one of the criteria that must be satisfied by a physically
meaningful definition of the current density.  Two other criteria are
important.  First, the formula must also reduce to the textbook
expression in regions of space that lie outside the range of the
nonlocal operators (pseudopotentials are typically confined to small
spheres surrounding the atoms).  Second, it must reduce to the
well-known expressions for the macroscopic current in the
long-wavelength limit.  The approaches that have been proposed so far
have either been specialized to a certain physical property (e.g.,
dielectric~\cite{Umari2001} or diamagnetic~\cite{Pickard2003}
response), or limited in scope to a subset of the above criteria. For
example, Li {\em et al.}~\cite{Li2008} proposed a strategy that
guarantees charge continuity by construction but does not satisfy the
two additional criteria, as we shall see in Sec.~\ref{curden}.

In addition to the technical challenges related to nonlocal
pseudopotentials, there is another complication associated with
the calculation of the flexoelectric coefficients using the current
density in bulk. Namely, the bulk nonlongitudinal responses contain
a contribution coming from the gradients of the local rotations in the
crystal.  This ``circulating rotation-gradient'' (CRG) contribution,
derived in Ref.~\onlinecite{StengelUNPUB} (where it is referred
to as a ``dynamic'' or ``gauge-field'' term),
must be treated carefully when comparing our calculations
with previous results. We will discuss this  point in Sec.~\ref{diamag}.

In this work we develop a first-principles methodology based on DFT to
calculate the full bulk CI FxE tensor from a single unit
cell.  At the heart of our technique lies the introduction of a
physically sound microscopic current-density operator in the presence of nonlocal
pseudopotentials
that fulfills all criteria that we
stated in the above paragraphs: (i) it satisfies the continuity
equation; (ii) the contribution of the nonlocal pseudopotentials is
correctly confined to the atomic spheres; and (iii) it reduces to the
macroscopic velocity operator in the long-wavelength limit.  We will
discuss our approach for calculating the current density in the
context of earlier works, and how it applies to the problem of
calculating bulk FxE coefficients. Finally, we will demonstrate
  that the results for the CI FxE coefficients from our current-density
  implementation are in excellent agreement with the previous
  charge-density-based DFT implementations described above
  \cite{Hong2013,Stengel2014}, confirming that it is an accurate and
  efficient method for calculating the FxE response of materials.

The paper is organized as follows. In Sec.~\ref{Approach} we outline
the general approach to determining FxE coefficients; in
Sec.~\ref{Form} we give the formalism used in our calculations of the
current density; in Sec.~\ref{Imp} we provide details of the
implementation of the formalism; Sec.~\ref{Res} presents benchmark
tests for the simple case of isolated noble gas atoms, and results for
several technologically important, cubic oxide compounds; in
Sec.~\ref{Disc}, we discuss some technical issues that are associated
with the current density in the presence of nonlocal pseudopotentials;
and we conclude the paper in Sec.~\ref{Con}.

\section{Approach\label{Approach}}

The goal of this work is to calculate the bulk CI
flexoelectric tensor elements
\begin{equation}
\label{muII}
\mu^{\text{I}}_{\alpha\beta,\omega\nu}=\frac{d
P_\alpha}{d\eta_{\beta,\omega\nu}},
\end{equation}
where $P_\alpha$ is the polarization in direction $\alpha$, and
\begin{equation}
\eta_{\beta,\omega\nu}=\frac{\partial^2u_\beta}{\partial
r_\omega\partial r_\nu}
\end{equation}
is the strain gradient tensor, where $u_\beta$ is the $\beta$
component of the displacement field. The superscript ``I'' indicates
that the tensor elements are defined with respect to the unsymmetrized
displacements \cite{Nye1985}; superscripts ``II'' will be used to
indicate tensor elements defined with respect to symmetrized strain.

Calculating the polarization in Eq.~(\ref{muII}) is tricky from a
quantum-mechanical standpoint, as it does not correspond to the
expectation value of a well-defined operator. As mentioned above, the
Berry-phase method \cite{KingSmith1993,Resta1994} can be used to
obtain the formal macroscopic polarization averaged over the
cell. However, we require access to the local polarization
\emph{density} $P_\alpha(\textbf{r})$.  Although the static
microscopic polarization density is not well defined in a quantum
mechanical context, at the linear-response level the \emph{induced}
polarization $P_{\alpha,\lambda}(\textbf{r})=\partial
P_\alpha(\textbf{r})/\partial\lambda$ resulting from a small change in
parameter $\lambda$ can be equated to the local current flow via
$\partial P_\alpha(\textbf{r})/\partial\lambda=
\partial J_\alpha(\textbf{r})/\partial\dot{\lambda}$, where
$\dot{\lambda}$ is the rate of change of the adiabatic parameter,
$\lambda$.  Following the approach of Ref.~\onlinecite{Stengel2013},
we now consider an adiabatic displacement of sublattice $\kappa$
(i.e., a given atom in the unit cell along with all of its
periodic images) of a
crystal in direction $\beta$ as given by
\begin{equation}
\label{phon}
u_{\kappa\beta}(l,t)=\lambda_{\kappa\beta\textbf{q}}(t)e^{i\textbf{q}\cdot\textbf{R}_{l\kappa}},
\end{equation}
where $l$ is the cell index. In this case, the induced local
polarization density $P_{\alpha,\kappa\beta\textbf{q}}(\textbf{r})$ in
direction $\alpha$ induced by mode $\kappa\beta$ of wavevector
$\textbf{q}$ is
\begin{equation}
\label{Jrt-dv}
P_{\alpha,\kappa\beta\textbf{q}}(\textbf{r})=
\frac{\partial J_{\alpha}(\textbf{r})}{\partial\dot{\lambda}_{\kappa\beta\textbf{q}}} .
\end{equation}
Using the fact that the linearly induced current will be modulated by
a phase with the same wavevector as the perturbation in
Eq.~(\ref{phon}), we can define
\begin{equation}
\label{Jrt}
P^{\textbf{q}}_{\alpha,\kappa\beta}(\textbf{r})=
P_{\alpha,\kappa\beta\textbf{q}}(\textbf{r})
e^{-i\textbf{q}\cdot\textbf{r}},
\end{equation}
which is therefore a lattice-periodic function.  This quantity, the
\emph{cell-periodic part of the first-order induced polarization
  density}, will play a central role in our considerations.  It is
also convenient to define
\begin{equation}
\label{Pbar}
\overline{P}_{\alpha,\kappa\beta}^{\textbf{q}} \equiv
\frac{1}{\Omega}\int_{\text{cell}}
P_{\alpha,\kappa\beta}^{\textbf{q}}(\textbf{r}) d^3r,
\end{equation}
where $\Omega$ is the cell volume, as the cell average of this
response.  In Ref.~\onlinecite{Stengel2013} it was shown that the
CI flexoelectric tensor elements are given by the second
wavevector derivatives of
$\overline{P}_{\alpha,\kappa\beta}^{\textbf{q}}$ via
\begin{equation}
\label{muI}
\begin{split}
\mu^{\text{I}}_{\alpha\beta,\omega\nu}&= -\frac{1}{2}\sum_\kappa\frac{\partial^2\overline{P}_{\alpha,\kappa\beta}^{\textbf{q}}}{\partial
q_\omega\partial q_\nu}\Bigg\vert_{\textbf{q}=0}.
\end{split}
\end{equation}

This formulation suggests that it may be possible to compute the
polarization responses
$\overline{P}^{\textbf{q}}_{\alpha,\kappa\beta}$ entirely from a
single-unit-cell calculation, similar to the way that phonon responses
are computed in DFPT.  In fact, this is the case. The formalism
necessary to compute these responses at the DFT level will be
presented in the next sections, giving access to an efficient and
robust means to compute the flexoelectric coefficients through
Eq.~(\ref{muI}).

\section{Formalism\label{Form}}

Given a time-dependent Hamiltonian with a single-particle solution
$\Psi(t)$, the current density at a point \textbf{r} in Cartesian
direction $\alpha$ can be written
\begin{equation}
  J_\alpha(\textbf{r})=
  \langle\Psi(t)\vert\hat{\mathcal{J}}_\alpha(\textbf{r})\vert\Psi(t)\rangle
\label{Js}
\end{equation}
where $\hat{\mathcal{J}}_\alpha(\textbf{r})$ is the current-density
operator (a caret symbol over a quantity will indicate an
operator). We will first address how to treat the time-dependent
wavefunctions (Sec.~\ref{adpert}), and then discuss the form of the
current-density operator in (Sec.~\ref{curden}) .

\subsection{Adiabatic density-functional perturbation theory\label{adpert}}

\subsubsection{Adiabatic response}

We write the time-dependent Schr\"odinger equation as 
\begin{equation}
\label{seq}
i\frac{\partial}{\partial t}\vert\Psi\rangle=\hat{H}(\lambda(t))\vert\Psi\rangle.
\end{equation}
where $\hat{H}(\lambda(t))$ is the Hamiltonian, and $\lambda$
parametrizes the time-dependent atomic motion.  Since we are
interested in the current density resulting from adiabatic
displacements, we expand the wavefunction $\vert\Psi(t)\rangle$ to first order in
the velocity, $\dot{\lambda}$: \cite{Messiah1981,Thouless1983,Niu1984}
\begin{equation}
\label{psiad}
 \vert\Psi(t)\rangle \simeq e^{i\gamma(t)}e^{i\phi(\lambda(t))}[\vert\psi(\lambda(t))\rangle+\dot{\lambda}(t)\vert\delta\psi(\lambda(t))\rangle],
\end{equation}
where $\vert\psi(\lambda)\rangle$ is the lowest-energy eigenfunction
of the time-independent Hamiltonian at a given $\lambda$, and
$\vert\delta\psi(\lambda)\rangle$ is the first order adiabatic
  wavefunction [defined by Eq.~(\ref{psiad})]; $\gamma(t)=-\int_0^t
E(\lambda(t^\prime))d t^\prime$ is the dynamic phase, with
  $E(\lambda)$ being the eigenenergy of $\vert\psi(\lambda)\rangle$;
$\phi(\lambda(t))=\int_0^t \langle\psi(\lambda(t^\prime))\vert
i\partial_t \psi(\lambda(t^\prime))\rangle d t^\prime$ is the
geometric Berry phase \cite{Berry1984} (we have used the shorthand
$\partial_t=\partial/\partial t$). We work in the parallel-transport
gauge, $\langle\psi(\lambda)\vert i\partial_\lambda
\psi(\lambda)\rangle=0$, so the Berry phase contribution vanishes.

Equation (\ref{psiad}) is written assuming a single occupied band, but
in the multiband case we shall let the evolution be guided by
multiband parallel transport instead.  In this case, the first-order
wavefunctions, $\delta\psi_n$, given by adiabatic perturbation
theory\cite{Messiah1981,Thouless1983,Niu1984}, are
\begin{equation}
\label{deltapsi}
\vert\delta\psi_n\rangle=-i\sum_{m}^\text{unocc}\vert\psi_m\rangle\frac{\langle\psi_m\vert\partial_\lambda\psi_n\rangle}{\epsilon_n-\epsilon_m},
\end{equation}
where $\epsilon_n$ is the eigenvalue of the $n$th single particle
wavefunction, and $\partial_\lambda$ is shorthand for
$\partial/\partial\lambda$.  The wavefunction
$\vert\partial_\lambda\psi_n\rangle$ is the first-order wavefunction
resulting from the \emph{static} perturbation
\begin{equation}
\label{delpsi}
\vert\partial_\lambda\psi_n\rangle=\sum_m^\text{unocc}\vert\psi_m\rangle\frac{\langle\psi_m\vert\partial_\lambda\hat{H}\vert\psi_n\rangle}{\epsilon_n-\epsilon_m},
\end{equation}
which is the quantity calculated in conventional DFPT implementations
\cite{Baroni2001,Gonze1997}.

\subsubsection{Density functional theory}

We will implement the calculations of the current density in the
context of plane-wave pseudopotential DFT, so the single-particle
wavefunctions we will use in Eq.~(\ref{deltapsi}) are solutions to the
Kohn-Sham equation for a given band $n$ and wavevector \textbf{k},
\begin{equation}
\label{KSeq}
\hat{H}_{\text{KS}}\vert\psi_{n\textbf{k}}\rangle=\epsilon_{n\textbf{k}}\vert\psi_{n\textbf{k}}\rangle.
\end{equation}
where the Kohn-Sham Hamiltonian is
\begin{equation}
\label{HKS}
\hat{H}_{\text{KS}}=\hat{T}_{\text{s}}+\hat{V}_{\text{H}}+\hat{V}_{\text{XC}}+\hat{V}_{\text{ext}}^{\text{loc}}+\hat{V}_{\text{ext}}^{\text{nl}}.
\end{equation}
Here $\hat{T}_{\text{s}}$ is the single-particle kinetic energy,
$\hat{V}_{\text{H}}$ is the Hartree potential, $\hat{V}_{\text{XC}}$
is the exchange correlation potential, and the external potential
contains both a local and nonlocal part (last two terms).  We will
consider norm-conserving, separable, Kleinmann-Bylander type
\cite{Kleinman1982} pseudopotentials. The form of the nonlocal
potential (henceforth referred to as $\hat{V}^{\text{nl}}$) is given
by Eq.~(\ref{VNL}). We will drop the ``KS'' subscript from here on.
Note that, although we focus on norm-conserving pseudopotentials
  in this work, the issues pertaining to nonlocal potentials that will
  be discussed in Sec.~\ref{curden} would apply to ultrasoft
  \cite{Vanderbilt1990} and projector augmented wave (PAW) \cite{Blochl1994}
  potentials as well.

\subsubsection{Polarization response}

Using the expansion in Eq.~(\ref{psiad}), the first-order one-particle
density matrix is
\begin{equation}
\label{denmat}
\delta\hat{\rho}=\dot{\lambda}\frac{2}{N_k}\sum_{n\textbf{k}}\left(\vert\delta\psi_{n\textbf{k}}\rangle\langle\psi_{n\textbf{k}}\vert+\vert\psi_{n\textbf{k}}\rangle\langle\delta\psi_{n\textbf{k}}\vert\right)
\end{equation}
where the factors $(2/N_k)\sum_{n\textbf{k}}$ take care of the spin
degeneracy, sum over occupied Bloch bands, and average over the
Brillouin zone.  A monochromatic perturbation such as that of
Eq.~(\ref{phon}) always comes together with its Hermitian conjugate,
coupling states at \textbf{k} with those at $\textbf{k}\pm\textbf{q}$,
so that each perturbed wavefunction has two components that we refer
as $\delta\psi_{n,\textbf{k}+\textbf{q}}$ and
$\delta\psi_{n,\textbf{k}-\textbf{q}}$ respectively.  We wish to
select the cross-gap response at $+\textbf{q}$, so we project onto
this component of the density matrix to obtain \cite{Adler1962}
\begin{equation}
\label{denmat2}
\delta\hat{\rho}_\textbf{q}=\dot{\lambda}\frac{2}{N_k}\sum_{n\textbf{k}}\left(
\vert\delta\psi_{n,\textbf{k}+\textbf{q}}\rangle\langle\psi_{n\textbf{k}}\vert
+
\vert\psi_{n\textbf{k}}\rangle\langle\delta\psi_{n,\textbf{k}-\textbf{q}}\vert
\right).
\end{equation}
Specializing now to the perturbation of Eq.~(\ref{phon}), the
corresponding polarization response is
\begin{equation}
\label{Plambda1}
\begin{split}
P_{\alpha,\kappa\beta\textbf{q}}(\textbf{r})
&=\frac{2}{N_k}\sum_{n\textbf{k}}\Big[\langle\psi_{n\textbf{k}}\vert\hat{\mathcal{J}}_\alpha(\textbf{r})\vert\delta\psi_{n\textbf{k},\textbf{q}}^{\kappa\beta}\rangle
\\
&\phantom{=\frac{2}{N_k}\sum_{n\textbf{k}}\Big[}+\langle\delta\psi_{n\textbf{k},-\textbf{q}}^{\kappa\beta}\vert \hat{\mathcal{J}}_\alpha(\textbf{r})\vert\psi_{n\textbf{k}}\rangle\Big].
\end{split}
\end{equation}
Using Eqs.~(\ref{deltapsi}) and (\ref{delpsi}), the needed
first-order wave functions are
\begin{equation}
\label{pertwf}
\vert\delta\psi^{\kappa\beta}_{n\textbf{k},\textbf{q}}\rangle=-i\sum_{m}^{\text{unocc}}\vert\psi_{m\textbf{k}+\textbf{q}}\rangle\frac{\langle\psi_{m\textbf{k}+\textbf{q}}\vert\partial_{\lambda_{\kappa\beta\textbf{q}}}\hat{H}\vert\psi_{n\textbf{k}}\rangle}{(\epsilon_{m\textbf{k}+\textbf{q}}-\epsilon_{n\textbf{k}})^2}.
\end{equation}

For Eq.~(\ref{muI}), we require the cell-average of the
$\textbf{q}$-dependent polarization response [Eq.~(\ref{Pbar})].
Defining the operator
\begin{equation}
\label{Jq0}
\hat{\mathcal{J}}_\alpha(\textbf{q})=\frac{1}{\Omega}
\int_{\rm cell} d^3r\,e^{-i\bf q\cdot r}\,\hat{\mathcal{J}}_\alpha(\textbf{r}),
\end{equation}
Eq.~(\ref{Pbar}) can be written
\begin{equation}
\begin{split}
\label{Pq}
  \overline{P}_{\alpha,\kappa\beta}^{\textbf{q}}&=\frac{2}{N_k}\sum_{n\textbf{k}} \Big[  \langle \psi_{n\textbf{k}}\vert\hat{\mathcal{J}}_\alpha(\textbf{q})\vert\delta \psi_{n\textbf{k},\textbf{q}}^{\kappa\beta}\rangle
\\
&\phantom{\frac{2}{N_k}\sum_{n\textbf{k}} \Big[}+\langle \delta \psi_{n\textbf{k},-\textbf{q}}^{\kappa\beta}\vert\hat{\mathcal{J}}_\alpha(\textbf{q})\vert \psi_{n\textbf{k}}\rangle \Big].
\end{split}
\end{equation}

The ground-state and first-order wavefunctions can be expressed in
terms of cell-periodic Bloch functions in the normal way:
\begin{equation}
\langle\textbf{s}\vert\psi_{n\textbf{k}}\rangle=u_{n\textbf{k}}(\textbf{s})e^{i\textbf{k}\cdot\textbf{s}}, \;\;\langle\textbf{s}\vert \delta \psi_{n\textbf{k},\textbf{q}}^{\kappa\beta}\rangle=\delta u^{\kappa\beta}_{n\textbf{k},\textbf{q}}(\textbf{s})e^{i(\textbf{k}+\textbf{q})\cdot\textbf{s}}.
\end{equation}
(Indices $\bf s$ and ${\bf s}'$ are not to be confused with the point
\textbf{r} at which the current density is evaluated.)  Using this
notation, the cell-periodic first-order static wavefunction is written
$\vert\partial_{\lambda}u^{\kappa\beta}_{n\textbf{k},\textbf{q}}\rangle$,
which is equivalent to $\vert
u_{n\textbf{k},\textbf{q}}^{\tau_{\kappa\beta}}\rangle$ in the
notation of Gonze and Lee \cite{Gonze1997} and $\vert \Delta
u_n^{\textbf{k}+\textbf{q}}\rangle$ in the notation of Baroni
\textit{et al.} \cite{Baroni2001}

By factoring out the phases with wavevector \textbf{k} and \textbf{q},
we can ensure that we only consider cell-periodic quantities, and
therefore all calculations can be performed on a unit
cell. \cite{Baroni2001} To this end, we define a cell-periodic
operator \footnote{Note that the definition of Eq.~(\ref{Jkqdef})
  involves a choice of convention in that the exponential factor
  $e^{i\textbf{q}\cdot\textbf{r}}$ is placed to the right of
  $\hat{\mathcal{J}}_\alpha(\textbf{q})$ as opposed to the
  left. Choosing the opposite convention would simply switch the
  operators between the two terms in Eq.~(\ref{Pq2}).}
\begin{equation}
\label{Jkqdef}
\hat{\mathcal{J}}_\alpha^{\textbf{k},\textbf{q}}=
  e^{-i\textbf{k}\cdot\hat{\textbf{r}}} \hat{\mathcal{J}}_\alpha(\textbf{q})
  e^{i(\textbf{k}+\textbf{q})\cdot\hat{\textbf{r}}} .
\end{equation}
Using the fact that $\hat{\mathcal{J}}_\alpha(\textbf{q})=
\hat{\mathcal{J}}^\dagger_\alpha(-\textbf{q})$ it follows that
$\left(\hat{\mathcal{J}}_\alpha^{\textbf{k},-\textbf{q}}\right)^\dagger=
e^{-i(\textbf{k}-\textbf{q})\cdot\hat{\textbf{r}}}
\hat{\mathcal{J}}_\alpha(\textbf{q})
e^{i\textbf{k}\cdot\hat{\textbf{r}}}$ so that Eq.~(\ref{Pq}) can be
written as
\begin{equation}
\begin{split}
\label{Pq2}
  \overline{P}_{\alpha,\kappa\beta}^{\textbf{q}}&=\frac{2}{N_k}\sum_{n\textbf{k}} \Big[  \langle u_{n\textbf{k}}\vert\hat{\mathcal{J}}_\alpha^{\textbf{k},\textbf{q}}\vert\delta u_{n\textbf{k},\textbf{q}}^{\kappa\beta}\rangle
\\
&\phantom{=\frac{2}{N_k}\sum_{n\textbf{k}} \Big[}+\langle \delta u_{n\textbf{k},-\textbf{q}}^{\kappa\beta}\vert\left(\hat{\mathcal{J}}_\alpha^{\textbf{k},-\textbf{q}}\right)^\dagger\vert u_{n\textbf{k}}\rangle \Big].
\end{split}
\end{equation}

In this work, we shall limit our focus to materials with time-reversal symmetry (TRS);
then we have
\begin{equation}
\label{TReq}
\langle \textbf{s} \vert u_{n\textbf{k}} \rangle= \langle u_{n-\textbf{k}}  \vert\textbf{s} \rangle, \;\;\langle\textbf{s}\vert \delta u_{n\textbf{k},\textbf{q}}^{\kappa\beta}\rangle=  -\langle \delta u_{n\,-\textbf{k},-\textbf{q}}^{\kappa\beta}\vert  \textbf{s}\rangle,
\end{equation}
where the negative sign in the second expression is a result of the
$-i$ in the first-order adiabatic wavefunction [see
Eq.~(\ref{deltapsi})]. Assuming that the current operator has the
correct ``TRS odd'' nature, i.e., $\Big( \langle {\bf
  s}|\hat{\mathcal{J}}^{\bf k,-q}_\alpha |{\bf s}' \rangle \Big)^* =
-\langle {\bf s}|\hat{\mathcal{J}}^{\bf -k,q}_\alpha |{\bf s}'
\rangle$, Eq.~(\ref{Pq2})
 simplifies to
\begin{equation}
\begin{split}
\label{PqTR}
  \overline{P}_{\alpha,\kappa\beta}^{\textbf{q}}
&=\frac{4}{N_k}\sum_{n\textbf{k}}  \langle u_{n\textbf{k}}\vert\hat{\mathcal{J}}_\alpha^{\textbf{k},\textbf{q}}
\vert\delta u_{n\textbf{k},\textbf{q}}^{\kappa\beta}\rangle.
\end{split}
\end{equation}

\subsection{Current-density operator\label{curden}}

We now consider the form of the current-density operator. If
  particle density is conserved, any physically meaningful definition
  of current density must satisfy the continuity condition
\begin{equation}
\label{conteq}
\nabla\cdot\textbf{J}(\textbf{r})=-\frac{\partial \rho(\textbf{r})}{\partial t},
\end{equation}
where $\rho$ is the particle density. In a quantum mechanical treatment\cite{Sakuri},
$\rho(\textbf{r})=\vert\Psi(\textbf{r})\vert^2$, where $\Psi$ is the
solution to the time-dependent Schr\"odinger equation. Combining
Eq.~(\ref{seq}) with its complex conjugate gives
\begin{equation}
\label{conteq2}
\frac{\partial}{\partial t}\rho(\textbf{r})=-i\langle\Psi\vert\left[\vert\textbf{r}\rangle\langle\textbf{r}\vert,\hat{H}\right]\vert\Psi\rangle=-i\langle\Psi\vert\left[\hat{\rho}(\textbf{r}),\hat{H}\right]\vert\Psi\rangle,
\end{equation}
where $\hat{\rho}(\textbf{r})$ is the particle density operator. (We
use atomic units throughout with an electron charge of $-1$.) In
  terms of the first-order adiabatic expansion of Eq.~(\ref{psiad}),
  we can use Eq.~(\ref{conteq2}) to write the induced density from an
  adiabatic perturbation parameterized by $\lambda$ as
\begin{equation}
\label{conteqlam}
\begin{split}
\rho_\lambda(\textbf{r})=-i\Big(&\langle\psi\vert\left[\hat{\rho}(\textbf{r}),\hat{H}\right]\vert\delta\psi\rangle+\langle\delta\psi\vert\left[\hat{\rho}(\textbf{r}),\hat{H}\right]\vert\psi\rangle\Big) .
\end{split}
\end{equation}

\subsubsection{Local potentials\label{locpot}}

Consider the simplest case of a Hamiltonian of the form
$\hat{H}^{\text{loc}}=\hat{\textbf{p}}^2/2 + \hat{V}^{\text{loc}}$
where $\hat{\textbf{p}}$ is the momentum operator and
$\hat{V}^{\text{loc}}=\int\hat{\rho}(\textbf{r})V(\textbf{r})d^3r$ is
a local scalar potential. The local potential commutes with the
density operator, so the only contribution to the current is from the
momentum operator. Comparing Eqs.~(\ref{conteq}) and (\ref{conteq2})
results in the textbook form of the current-density operator
\begin{equation}
\label{jloc}
\begin{split}
\hat{\mathcal{J}}_\alpha^{\text{loc}}(\textbf{r})&=-\frac{1}{2}\left(\vert\textbf{r}\rangle\langle\textbf{r}\vert\hat{p}_\alpha+\hat{p}_\alpha\vert\textbf{r}\rangle\langle\textbf{r}\vert\right)
\\
&=-\frac{1}{2}\left\{\hat{\rho}(\textbf{r}),\hat{p}_\alpha\right\}.
\end{split}
\end{equation}
Using Eq.~(\ref{Jq0}), we have
\begin{equation}
\label{jqloc}
\begin{split}
\hat{\mathcal{J}}_\alpha^{\text{loc}}(\textbf{q})&=-\frac{1}{2}\left(e^{-i\textbf{q}\cdot\hat{\textbf{r}}}\hat{p}_\alpha+\hat{p}_\alpha e^{-i\textbf{q}\cdot\hat{\textbf{r}}}\right),
\end{split}
\end{equation}
which gives the cell-periodic operator (Appendices \ref{sepICL} and \ref{Jloc})
\begin{equation}
\label{jkqloc}
\begin{split}
\hat{\mathcal{J}}_\alpha^{\textbf{k},\textbf{q},\text{loc}}&=-\left(\hat{p}_\alpha^\textbf{k}+\frac{q_\alpha}{2}\right),
\end{split}
\end{equation}
where
$\hat{p}_\alpha^{\textbf{k}}=-i\hat{\nabla}_\alpha+\hat{k}_\alpha$ is
the cell-periodic momentum operator ($\hat{\nabla}_\alpha$ is a
spatial derivative in the $\alpha$ direction, and the overall minus
sign is from the electron charge).

\subsubsection{Continuity condition and nonlocal potentials\label{contsec}}

As mentioned above, \emph{nonlocal} potentials are ubiquitous in
modern pseudopotential implementations of DFT
\cite{Vanderbilt1990,Hamann1979,Kleinman1982,Blochl1994}. When
nonlocal potentials are present in the Hamiltonian, the current
density in Eq.~(\ref{jloc}) does not satisfy the continuity equation.

To see this, consider a Hamiltonian with a nonlocal potential:
$\hat{H}^{\text{nl}}=\hat{\textbf{p}}^2/2 +
\hat{V}^{\text{loc}}+\hat{V}^{\text{nl}}$ with
$\hat{V}^{\text{nl}}=\int d^3r\int d^3r^\prime
\hat{\rho}(\textbf{r},\textbf{r}^\prime)V(\textbf{r},\textbf{r}^\prime)$
where
$\hat{\rho}(\textbf{r},\textbf{r}^\prime)=\vert\textbf{r}\rangle\langle\textbf{r}^\prime\vert$.
In this case, there is a term in the induced density [Eq.~(\ref{conteqlam})] resulting from the nonlocal potential:
\begin{equation}
\begin{split}
\label{rhoNL}
\rho^{\text{nl}}_\lambda(\textbf{r})=-i\Big(&\langle\psi\vert\left[\hat{\rho}(\textbf{r}),\hat{V}^{\text{nl}}\right]\vert\delta\psi\rangle
\\
&+\langle\delta\psi\vert\left[\hat{\rho}(\textbf{r}),\hat{V}^{\text{nl}}\right]\vert\psi\rangle\Big),
\end{split}
\end{equation}
If we write the total induced
current as the sum of contributions from the local and nonlocal parts,
$\textbf{J}=\textbf{J}^{\text{loc}}+\textbf{J}^{\text{nl}}$, then we
have
\begin{equation}
\label{conteq3}
\nabla\cdot\textbf{J}^{\text{nl}}(\textbf{r})=-\rho^{\text{nl}}_\lambda(\textbf{r}).
\end{equation}
This ``nonlocal charge,'' $\rho^{\text{nl}}_\lambda$, measures the
degree to which the continuity equation, Eq.~(\ref{conteq}), breaks
down if Eq.~(\ref{jloc}) is used in a nonlocal pseudopotential
context.

Li \textit{et al.}\cite{Li2008} argued that such nonlocal charge could
be used to reconstruct the nonlocal contribution to the current
density via a Poisson equation. Indeed, Eq.~(\ref{conteq3}) indicates
that the irrotational part of $\textbf{J}^{\text{nl}}$ can be
determined by calculating Eq.~(\ref{rhoNL}). Their approach yields a
conserved current by construction, but there are two additional
requirements that a physically meaningful definition of the
quantum-mechanical electronic current should satisfy:
\begin{itemize}
\item The nonlocality of the Hamiltonian should be confined to small
  spheres surrounding the ionic cores. In the interstitial regions,
  the nonlocal part of the pseudopotentials vanish, and the
  Hamiltonian operator is local therein. Thus, the current-density
  operator should reduce to the simple textbook formula outside the
  atomic spheres. The corollary is that
  $\textbf{J}^{\text{nl}}(\textbf{r})$ must vanish in the interstitial
  regions.

\item The macroscopic average of the microscopic current should reduce
  to the well-known expression
  $\hat{v}_\alpha= -i[\hat{r}_\alpha,\hat{H}]$ for the electronic
  velocity operator
  \cite{Starace1971,Hybertsen1987,Giannozzi1991,DalCorso1994}. This is
  routinely used in the context of DFPT, e.g., to calculate the
  polarization response to ionic displacements needed for the Born
  effective charge tensor.
\end{itemize}

The strategy proposed by Li {\em et al.}~\cite{Li2008} falls short of
fulfilling either condition. Regarding the first (spatial
confinement), note that the nonlocal charge associated to individual
spheres generally has a nonzero dipole (and higher multipole)
moments. Therefore, even if the nonlocal charge is confined to the
sphere, an irrotational field whose divergence results in such a
charge density will generally have a long-ranged character and
propagate over all space.

Regarding the relation to the macroscopic particle velocity, note that
the construction proposed by Li {\em et al.}~\cite{Li2008} in practice
discards the solenoidal part of the nonlocal current and hence fails
at describing its contribution to the transverse polarization
response. This is precisely the quantity in which we are interested in the
context of flexoelectricity, and is also crucial for obtaining other
important quantities, such as the Born charge tensor, that are part of
standard DFPT implementations.

Therefore, a calculation of Eqs.~(\ref{rhoNL}) does not contain the
necessary information to determine $\textbf{J}^{\text{nl}}$, and an
alternative derivation to the textbook one outlined in
Sec.~\ref{locpot} is required.

\subsubsection{Current-density operator  generalized for nonlocal potentials\label{secJ}}

In light of the previous section, we will now focus on determining an
expression for $\hat{\mathcal{J}}_\alpha$ that is applicable when
nonlocal potentials are present in the Hamiltonian. For the case of a
perturbation that is uniform over the crystal, corresponding to the
long wavelength $\textbf{q}=0$ limit of Eq.~(\ref{phon}), it is well
known that the momentum operator should be replaced with the canonical
velocity operator $\hat{v}_\alpha$
\cite{Starace1971,Hybertsen1987,Giannozzi1991,DalCorso1994} in order
to determine the \emph{macroscopic} current.

In Ref.~\onlinecite{Umari2001}, the expression for the
\emph{microscopic} current operator that was used to calculate the
current induced by a uniform electric field was Eq.~(\ref{jloc}) with
$\hat{p}_\alpha$ replaced by $\hat{v}_\alpha$.  Although this treatment
will result in the correct current when averaged over a unit cell,
this operator does not satisfy the continuity condition in
Eq.~(\ref{conteq}) except in the special case of a Hamiltonian with
only local potentials, where it reduces to Eq.~(\ref{jloc}).

Since we shall be treating a long wavelength acoustic phonon in this study,
and we require the polarization response be correct at least to second
order in \textbf{q} [\textit{cf.} Eq.~(\ref{muI})], we
require a version of $\hat{\mathcal{J}}_\alpha$ that is designed to handle spatially
varying perturbations.
Therefore, for our purposes, we need an alternative starting
  point for the derivation of a current-density expression,
  different from the one
  based on the continuity condition that led to, e.g.,
  Eq.~(\ref{jloc}).

In general, for an arbitrary electronic Hamiltonian $\hat{H}^{\textbf{A}}$ coupled to a
vector potential $\textbf{A}(\textbf{r})$, the most general form for
the current-density operator is
\begin{equation}
\label{dHdA}
\hat{\mathcal{J}}_\alpha(\textbf{r})=-\frac{\partial\hat{H}^{\textbf{A}}}{\partial A_\alpha(\textbf{r})} .
\end{equation}
Our strategy will be to use a vector potential to probe the response to the strain gradient, which will give us the current density via Eq.~(\ref{dHdA}). 
Since we are treating the strain gradient in terms of a long-wavelength acoustic phonon of
wavevector $\bf q$, and we are interested in the response occurring at
the same wavevector $\bf q$,
it is useful to define
\begin{align}
\hat{\mathcal{J}}_\alpha(\textbf{r})&=\sum_{\textbf{G}}\hat{\mathcal{J}}_\alpha(\textbf{G}+\textbf{q})e^{i(\textbf{G}+\textbf{q})\cdot\textbf{r}},
\label{fft-J}
\\
A_\alpha(\textbf{r})&=\sum_{\textbf{G}}A_\alpha(\textbf{G}+\textbf{q})e^{i(\textbf{G}+\textbf{q})\cdot\textbf{r}},
\label{fft-A}
\\
P_{\alpha,\kappa\beta\textbf{q}}(\textbf{r})&=\sum_{\textbf{G}}
P_{\alpha,\kappa\beta\textbf{q}}(\textbf{G}+\textbf{q})e^{i(\textbf{G}+\textbf{q})\cdot\textbf{r}}.
\label{fft-dlP}
\end{align}
With these definitions, Eq.~(\ref{dHdA}) becomes
\begin{equation}
\label{dHdAGq}
\hat{\mathcal{J}}_\alpha(\textbf{G}+\textbf{q})=-\frac{\partial\hat{H}^{\textbf{A}}}{\partial A^*_\alpha(\textbf{G}+\textbf{q})}
\end{equation}
and the desired operator for Eq.~(\ref{Pq}) is
\begin{equation}
\label{dHdAq}
\hat{\mathcal{J}}_\alpha({\textbf{q}})=
-\frac{\partial\hat{H}^{\textbf{A}}}{\partial A^*_\alpha(\textbf{q})}.
\end{equation}

Again, if the Hamiltonian of interest had the form of
$H^{\text{loc}}=(\hat{\textbf{p}}+\hat{\textbf{A}})^2/2 +
\hat{V}^{\text{loc}}$,
where the scalar potential is local and
$\hat{\textbf{A}}=\int\hat{\rho}(\textbf{r})\textbf{A}(\textbf{r})d^3r$
is a local vector potential, then
$\hat{\mathcal{J}}^{\text{loc}}_\alpha(\textbf{r})=-\frac{1}{2}
\left\{\hat{\rho}(\textbf{r}),(\hat{p}_\alpha+\hat{A}_\alpha)\right\}$.
However, for our implementation, we are considering the case where the
potential $\hat{V}$ is nonlocal, so we must determine how to couple a
generally nonlocal Hamiltonian to a spatially nonuniform vector
potential field (which will be the case for a finite \textbf{q}
perturbation).

The standard strategy for describing the coupling to the vector
potential is to multiply the nonlocal operator by a complex phase
containing the line integral of the vector potential
\textbf{A}\cite{ICL2001,Pickard2003,Essin2010}; in the real-space
representation:
\begin{equation}
\label{Aphase}
\mathcal{O}^{\textbf{A}}(\textbf{s},\textbf{s}^\prime)=\mathcal{O}(\textbf{s},\textbf{s}^\prime)e^{-i\int_{\textbf{s}^\prime\rightarrow\textbf{s}}\textbf{A}\cdot d\ell}.
\end{equation}
The different methods that have been proposed for coupling \textbf{A}
to a nonlocal Hamiltonian amount to applying the complex phase in
Eq.~(\ref{Aphase}) to either the entire Hamiltonian\cite{Essin2010} or
just the nonlocal potential\cite{ICL2001,Pickard2003}, and choosing
either a straight-line path\cite{ICL2001,Essin2010} or a path that
passes through the centers of the atoms\cite{Pickard2003} to perform
the line integral.

\subsubsection{Straight-line path \label{formICL}}
Using Feynman path integrals, Ismail-Beigi, Chang, and Louie
\cite{ICL2001} (ICL) derived the following form of a nonlocal
Hamiltonian coupled to a vector potential field:
\begin{equation}
\label{HICL}
\begin{split}
\hat{H}^{\textbf{A}}_{\text{ICL}}&=\frac{1}{2}(\hat{\textbf{p}}+\hat{\textbf{A}})^2+\hat{V}^{\text{loc}}
\\
&+\int d^3s \int d^3s^\prime\hat{\rho}(\textbf{s},\textbf{s}^\prime)V^{\text{nl}}(\textbf{s},\textbf{s}^\prime)e^{-i\int_{\textbf{s}^\prime}^{\textbf{s}}\textbf{A}\cdot d\ell},
\end{split}
\end{equation}
where the line integral is taken along a straight path from \textbf{s}
to $\textbf{s}^\prime$.  
Since the approach used in Ref.~\onlinecite{ICL2001} to perform
  the minimal substitution
  $\hat{\textbf{p}}\rightarrow\hat{\textbf{p}}+\hat{\textbf{A}}$ is
  general, applying to both local and nonlocal Hamiltonians, this
  approach is equivalent to the
approach of Essin \textit{et al.}, where the coupled Hamiltonian is
written as
\begin{equation}
\label{HA1}
H^{\textbf{A}}(\textbf{s},\textbf{s}^\prime)=H(\textbf{s},\textbf{s}^\prime)e^{-i\int_{\textbf{s}^\prime}^{\textbf{s}}\textbf{A}\cdot d\ell},
\end{equation}
i.e., all of the \textbf{A} dependence is contained in the complex
phase, and the line integral is also taken along a straight path from
\textbf{s} to $\textbf{s}^\prime$.

Expanding Eq.~(\ref{HA1}) to first order gives
\begin{equation}
\label{HA2}
\begin{split}
  H^{\textbf{A}}(\textbf{s},\textbf{s}^\prime)&= H(\textbf{s},\textbf{s}^\prime)-iH(\textbf{s},\textbf{s}^\prime)\int_{\textbf{s}^\prime}^{\textbf{s}}\textbf{A}\cdot d\ell+\cdots.
\end{split}
\end{equation}
We would like to evaluate Eq.~(\ref{dHdAq}) for this form of the
Hamiltonian.  Since $\textbf{A}(\textbf{r})$ is real we can write
Eq.~(\ref{fft-A}) as $A_\alpha(\textbf{r})=A^*_\alpha(\textbf{r})
=A^*_\alpha(\textbf{q})e^{-i\textbf{q}\cdot\textbf{r}}$ so that
 the integral over \textbf{A} for the ICL\cite{ICL2001}
path is
\begin{equation}
\label{AICL}
\begin{split}
\int_{\textbf{s}^\prime}^{\textbf{s}}\textbf{A}\cdot d\ell&=\int_0^1d\tau\textbf{A}[\textbf{s}^\prime+\tau(\textbf{s}-\textbf{s}^\prime)]\cdot(\textbf{s}-\textbf{s}^\prime)
\\
&=\textbf{A}^*(\textbf{q})\cdot(\textbf{s}-\textbf{s}^\prime)\int_0^1d\tau e^{-i{\textbf{q}}\cdot[\textbf{s}^\prime+\tau(\textbf{s}-\textbf{s}^\prime)]}
\\
&=-\textbf{A}^*(\textbf{q})\cdot(\textbf{s}-\textbf{s}^\prime)\frac{e^{-i\textbf{q}\cdot\textbf{s}}-e^{-i\textbf{q}\cdot\textbf{s}^\prime}}{i\textbf{q}\cdot(\textbf{s}-\textbf{s}^\prime)}
\end{split}
\end{equation}
Therefore, from Eqs.~(\ref{HA2}) and (\ref{dHdAq}),
\begin{equation}
\label{JqSL}
\langle\textbf{s}\vert\hat{\mathcal{J}}_{\alpha}^{\text{ICL}}(\textbf{q})\vert\textbf{s}^\prime\rangle=-iH(\textbf{s},\textbf{s}^\prime)(s_\alpha-s_\alpha^\prime)\frac{e^{-i\textbf{q}\cdot\textbf{s}}-e^{-i\textbf{q}\cdot\textbf{s}^{\prime}}}{i\textbf{q}\cdot(\textbf{s}-\textbf{s}^\prime)}.
\end{equation}
In practice we shall normally work in terms of the cell-periodic
current operator of Eq.~(\ref{Jkqdef}), whose position representation
follows as
\begin{equation}
\label{JkqICL}
\langle\textbf{s}\vert\hat{\mathcal{J}}_{\alpha}^{\textbf{k},\textbf{q},\text{ICL}}\vert\textbf{s}^\prime\rangle=-iH^{\textbf{k}}(\textbf{s},\textbf{s}^\prime)(s_\alpha-s_\alpha^\prime)\frac{e^{-i\textbf{q}\cdot(\textbf{s}-\textbf{s}^\prime)}-1}{i\textbf{q}\cdot(\textbf{s}-\textbf{s}^\prime)}.
\end{equation}

We can see that the current operator of Eq.~(\ref{JqSL}) satisfies the
continuity condition of Eq.~(\ref{conteq}) as follows. In reciprocal
space the continuity equation becomes
$i\textbf{q}\cdot[-\hat{\mathcal{J}}^{\text{ICL}}(\textbf{q})]=
-\partial\hat{\rho}_{\textbf{q}}/\partial t$, where
$\hat{\rho}_{\textbf{q}}=e^{-i\textbf{q}\cdot\hat{\textbf{r}}}$ is the
$\textbf{G}=0$ particle density operator for a given \textbf{q}, and
the negative sign in front of the current operator reflects the sign
of the electron charge. But from Eq.~(\ref{JqSL}) it quickly follows
that
\begin{equation}
\label{JqSLp}
-i\textbf{q}\cdot\langle\textbf{s}\vert\hat{\mathcal{J}}_{\alpha}^{\text{ICL}}(\textbf{q})\vert\textbf{s}^\prime\rangle=
i\langle\textbf{s}\vert\left[\hat{\rho}_\textbf{q},\hat{H}\right]\vert\textbf{s}^\prime\rangle
\end{equation}
which, using the Ehrenfest theorem, is nothing other than $-\partial
\hat{\rho}_{\textbf{q}}/\partial t$ in the position representation.

In the case that only local potentials are present, only the kinetic
term in the Hamiltonian contributes to
$\hat{\mathcal{J}}_{\alpha}^{\text{ICL}}(\textbf{q})$. We show in
Appendix \ref{sepICL} that the current operator then reduces to the
form of Eq.~(\ref{jqloc}).  The fact that the local and nonlocal parts
can be separated confirms the equivalence of the ICL
[Eq.~(\ref{HICL})] and Essin \textit{et al.} [Eq.~(\ref{HA1})]
approaches.

In the case that nonlocal potentials are present, we show in Appendix
\ref{sepICL} that, for $\textbf{q}=0$, Eq.~(\ref{JqSL}) reduces to the
well-known expression for the canonical velocity
operator\cite{Starace1971,Hybertsen1987,Giannozzi1991,DalCorso1994}
$\hat{\mathcal{J}}^{\text{ICL}}_{\alpha}(\textbf{q}=0)=-\hat{v}_{\alpha}=i\left[\hat{r}_\alpha,\hat{H}\right]$,
where the $-1$ comes from the electron charge. We discuss the case of
nonlocal potentials and finite \textbf{q} perturbations in
Sec.~\ref{longwave}.

\subsubsection{Path through atom center\label{formPM}}

Subsequently, Pickard and Mauri \cite{Pickard2003} (PM) proposed using
a path from \textbf{s} to the atom center, \textbf{R}, and then to
$\textbf{s}^\prime$, which was constructed explicitly to give better
agreement for magnetic susceptibility between pseudopotential and
all-electron calculations. This approach can be regarded as a
  generalization to spatially nonuniform fields of the gauge-including
  projector augmented-wave (GIPAW) method
  \cite{Pickard2001,Pickard2003}, where the PAW transformation is
  modified with a complex phase in order to ensure that the
  pseudowavefunction has the correct magnetic translational symmetry.

The coupled
Hamiltonian used in Ref.~\onlinecite{Pickard2003} is of the form
\begin{equation}
\label{HPM}
\begin{split}
\hat{H}^{\textbf{A}}_{\text{PM}}&=\frac{1}{2}(\hat{\textbf{p}}+\hat{\textbf{A}})^2+\hat{V}^{\text{loc}}+\sum_{\zeta=1}^N\int d^3s \int d^3s^\prime
\\
&\times \hat{\rho}(\textbf{s},\textbf{s}^\prime)V^{\text{nl}}_\zeta(\textbf{s},\textbf{s}^\prime)e^{-i\int_{\textbf{s}^\prime\rightarrow\textbf{R}_\zeta\rightarrow\textbf{s}}\textbf{A}\cdot d\ell},
\end{split}
\end{equation}
where $N$ is the number of atoms in the cell, $\textbf{R}_\zeta$ is
the position of atom  
$\zeta$, and $V_\zeta^{\text{nl}}$ is the nonlocal
potential for that atom. The PM approach explicitly splits the
nonlocal contribution from \textbf{A} into contributions from each
atomic sphere centered at $\textbf{R}_\zeta$.~\footnote{In contrast to
  the ICL straight-line path, Eq.~(\ref{HA1}) using the PM
  $\textbf{s}^\prime\rightarrow\textbf{R}_\zeta\rightarrow\textbf{s}$
  path [i.e., the phase in Eq.~(\ref{HPM}) multiplying the entire
  Hamiltonian instead of just
  $V^{\text{nl}}(\textbf{s},\textbf{s}^\prime)$] does \emph{not}
  recover $\hat{\mathcal{J}}_\alpha^{\text{loc}}$ for local
  potentials.} Therefore, the total current operator is
\begin{equation}
\begin{split}
\label{JkqPM}
\hat{\mathcal{J}}_{\alpha}^{\textbf{k},\textbf{q},\text{PM}}=-\left(\hat{p}^{\textbf{k}}_\alpha+\frac{q_\alpha}{2}\right)+\sum_{\zeta=1}^N \hat{\mathcal{J}}_{\alpha,\zeta}^{\textbf{k},\textbf{q},\text{PM,nl}},
\end{split}
\end{equation}
where the superscript ``nl'' and the subscript $\zeta$ emphasize that
each item in the summation describes the contribution to the current
from the nonlocal potential of the atom $\zeta$; it is obvious from
Eqs.~(\ref{HPM}) and (\ref{JkqPM}) that
$\hat{\mathcal{J}}_\alpha^{\text{loc}}$ will be recovered in the case
of a local potential.

\begin{widetext}
For an atom at position $\textbf{R}_\zeta$, the line integral in Eq.~(\ref{HPM}) is
\begin{equation}
\begin{split}
\int_{\textbf{s}^\prime\rightarrow\textbf{R}_\zeta \rightarrow\textbf{s}}\textbf{A}\cdot d\ell&=-\textbf{A}^*(\textbf{q})\cdot(\textbf{R}_\zeta -\textbf{s}^\prime)\frac{e^{-i\textbf{q}\cdot\textbf{R}_\zeta }-e^{-i\textbf{q}\cdot\textbf{s}^\prime}}{i\textbf{q}\cdot(\textbf{R}_\zeta -\textbf{s}^\prime)}-\textbf{A}^*(\textbf{q})\cdot(\textbf{s}-\textbf{R}_\zeta )\frac{e^{-i\textbf{q}\cdot\textbf{s}}-e^{-i\textbf{q}\cdot\textbf{R}_\zeta }}{i\textbf{q}\cdot(\textbf{s}-\textbf{R}_\zeta )}.
\end{split}
\end{equation}
 Therefore we have
\begin{equation}
\label{JqPMNL}
\begin{split}
\langle\textbf{s}\vert\hat{\mathcal{J}}_{\alpha,\zeta}^{\text{PM},\text{nl}}(\textbf{q})\vert\textbf{s}^\prime\rangle&=-iV_\zeta^{\text{nl}}(\textbf{s},\textbf{s}^\prime)\bigg[(R_{\alpha,\zeta}-s^\prime_\alpha)\frac{e^{-i\textbf{q}\cdot\textbf{R}_\zeta }-e^{-i\textbf{q}\cdot\textbf{s}^\prime}}{i\textbf{q}\cdot(\textbf{R}_\zeta -\textbf{s}^\prime)}+(s_\alpha-R_{\alpha,\zeta})\frac{e^{-i\textbf{q}\cdot\textbf{s}}-e^{-i\textbf{q}\cdot\textbf{R}_\zeta }}{i\textbf{q}\cdot(\textbf{s}-\textbf{R}_\zeta )}\bigg],
\end{split}
\end{equation}
so the cell-periodic operator is
\begin{equation}
\begin{split}
\label{JkqPMNL}
\langle\textbf{s}\vert\hat{\mathcal{J}}_{\alpha,\zeta}^{\textbf{k},\textbf{q},\text{PM,nl}}\vert\textbf{s}^\prime\rangle=-iV^{\text{nl}}_\zeta(\textbf{s},\textbf{s}^\prime)&\bigg[(R_{\alpha,\zeta}- s^\prime_\alpha)\frac{e^{-i\textbf{q}\cdot(\textbf{R}_\zeta-\textbf{s}^\prime)}-1}{i\textbf{q}\cdot(\textbf{R}_\zeta-\textbf{s}^\prime)}
+(s_\alpha-R_{\alpha,\zeta})\frac{e^{-i\textbf{q}\cdot(\textbf{s}-\textbf{s}^\prime)}-e^{-i\textbf{q}\cdot(\textbf{R}_\zeta-\textbf{s}^\prime)}}{i\textbf{q}\cdot(\textbf{s}-\textbf{R}_\zeta)}\bigg].
\end{split}
\end{equation}
\end{widetext}

From Eqs.~(\ref{JqPMNL}) and (\ref{rhoNL}), we see that
$i\textbf{q}\cdot[-\hat{\mathcal{J}}^{\text{PM},\text{nl}}(\textbf{q})]=i\left[e^{-i\textbf{q}\cdot\hat{\textbf{r}}},\hat{V^{\text{nl}}}\right]=-\hat{\rho}_{\lambda}^{\text{nl}}$. Therefore,
Eq.~(\ref{JkqPM}) satisfies the continuity condition. Also, in the
case of a $\textbf{q}=0$ perturbation,
$\hat{\mathcal{J}}^{\text{PM},\text{nl}}_{\alpha}(\textbf{q}=0)=i\left[\hat{r}_\alpha,\hat{V}^{\text{nl}}\right]$,
which is the nonlocal contribution to $-\hat{v}_\alpha$, as
expected. We discuss the case of nonlocal potentials and finite
\textbf{q} perturbations in the next section.

Finally, we see that for the longitudinal response (where
$\textbf{q}=q_\alpha\hat{\alpha}$), the ICL and PM approaches produce
identical operators. This is expected, since they both satisfy the
continuity equation. Only circulating currents (e.g., transverse or
shear FxE components) may exhibit path dependence.

\subsection{Long wavelength expansion \label{longwave}}

Recall that only the induced polarization up to second order in
$\textbf{q}$ is required for the FxE coefficients
[\textit{cf.}~Eq.~(\ref{muI})]. Therefore, instead of attempting to
calculate Eq.~(\ref{PqTR}) with either Eq.~(\mbox{\ref{JkqICL}}) or
(\ref{JkqPM}) directly, we will expand these expressions for the
current-density operator to second order in \textbf{q}.

Considering the Hamiltonian in Eq.~(\ref{HKS}), there are
contributions to $\hat{\mathcal{J}}_\alpha^{\textbf{q}}$ from the
kinetic energy and nonlocal part of the pseudopotential. We show in
Appendix \ref{sepICL} [Eq.~(\ref{LocNL4})] that the kinetic energy
only contributes up to first order in \textbf{q}, and for a local
Hamiltonian, the current operator reduces to the form of
Eq.~(\ref{jkqloc}).

The nonlocal potential will, however, contribute at all orders. As
mentioned in Sec.~\ref{formICL} and \ref{formPM}, for $\textbf{q}=0$,
both the ICL and PM approaches give
$\hat{\mathcal{J}}_\alpha^{\textbf{k},\textbf{q}=0}=-\hat{v}_\alpha^{\textbf{k}}=i[\hat{r}_\alpha,\hat{H}^{\textbf{k}}]=-\hat{p}_\alpha^{\textbf{k}}+\hat{\mathcal{J}}_\alpha^{\textbf{k},\text{nl}(0)}$,
where we have defined
$\hat{\mathcal{J}}_\alpha^{\textbf{k},\text{nl}(0)}\equiv
i[\hat{r}_\alpha,\hat{V}^{\textbf{k}\text{,nl}}]$. At higher orders in
\textbf{q}, and for nonlongitudinal response, the ICL and PM
approaches may no longer agree.

Up to second order in \textbf{q}, the current operator can be written as
\begin{equation}
\begin{split}
\label{JqExpand} \hat{\mathcal{J}}_{\alpha}^{\textbf{k},\textbf{q}}&\simeq-\left(\hat{p}_\alpha^{\textbf{k}}+\frac{q_\alpha}{2}\right) +\hat{\mathcal{J}}_\alpha^{\textbf{k},\text{nl}(0)}
\\
&\phantom{=}+ \frac{q_\gamma}{2} \hat{\mathcal{J}}_{\alpha,\gamma}^{\textbf{k},\text{nl}(1)} +\frac{q_\gamma q_\xi}{6}\hat{\mathcal{J}}_{\alpha,\gamma \xi}^{\textbf{k},\text{nl}(2)}.
\end{split}
\end{equation}
where the higher order terms in \textbf{q} ($\hat{\mathcal{J}}_{\alpha,\gamma}^{\textbf{k},\textbf{q},\text{nl}(1)}$
and
$\hat{\mathcal{J}}_{\alpha,\gamma\xi}^{\textbf{k},\textbf{q},\text{nl}(2)}$)
are the result of the nonlocal part of the Hamiltonian \emph{and} the
fact that the monochromatic perturbation is nonuniform (i.e, finite
\textbf{q}). Expressions for these last two terms in
  Eq.~(\ref{JqExpand}) are derived in Appendix \ref{JNL} for the ICL
  path [Eqs.~(\ref{ICL1}) and (\ref{ICL2})] and PM path
  [Eqs.~(\ref{PM1}) and (\ref{PM2})].

Plugging the current operator from Eq.~(\ref{JqExpand})
into Eq.~(\ref{PqTR}), readily yields the induced polarization,
\begin{equation}
\overline{P}_{\alpha,\kappa\beta}^{\textbf{q}} =  \overline{P}^{\textbf{q},\text{loc}}_{\alpha,\kappa\beta} + \overline{P}^{\textbf{q},\text{nl}}_{\alpha,\kappa\beta},
\label{PqExpand}
\end{equation}
where we have separated the contribution of the local current operator
(loc) from the nonlocal (nl) part.  The exact expression for
$\overline{P}^{\textbf{q},\text{loc}}_{\alpha,\kappa\beta}$ is derived
in Appendix \ref{Jloc}, yielding Eq.~(\ref{pkq}); the approximate
(exact only up to second order in ${\bf q}$) expression for
$\overline{P}^{\textbf{q},\text{nl}}_{\alpha,\kappa\beta}$ is derived
in Appendix \ref{JNL} [see Eq.~(\ref{Pqexpand2})].

\subsection{Circulating rotation-gradient contribution and diamagnetic susceptibility \label{diamag} }

Transverse or shear strain gradients result in rigid rotations of unit
cells which must be treated carefully in order to calculate physically
meaningful values of the flexoelectric tensor. This issue can be
loosely compared to the well-known distinction between the proper and
improper piezoelectric tensor,~\cite{Martin1972,Vanderbilt2000} but,
in the case of strain gradients, it is complicated by the fact that
different parts of the sample typically rotate by different amounts.
The reader is referred to Ref.~\onlinecite{StengelUNPUB} for a complete
discussion; only the results of that work necessary for our purposes
will be reproduced here.

Larmor's theorem states that the effects of a uniform rotation and
those of a uniform magnetic field are the same to first order in the
field/angular velocity. Therefore, the local rotations of the sample
dynamically produce circulating diamagnetic currents that will
contribute to the bulk flexoelectric coefficients as defined in
Eq.~(\ref{muI}). As was shown in Ref.~\onlinecite{StengelUNPUB} (see
also Appendix \ref{Appdiamag} for an abridged derivation), this
circulating rotation-gradient (CRG) \footnote{Recall that in
  Ref.~\onlinecite{StengelUNPUB}, this contribution is referred to as
  the ``dynamic'' gauge-field term.}  contribution only concerns the
nonlongitudinal components and is proportional to the diamagnetic
susceptibility of the material, $\chi_{\gamma\lambda}=\partial
M_\gamma/\partial H_\lambda$, where $M$ is the magnetization and $H$
the magnetic field. Specifically,
\begin{equation}
\label{pchi}
\begin{split}
\overline{P}^{(2,\omega\nu),\text{CRG}}_{\alpha,\beta}&=\sum_{\gamma\lambda}\left(\epsilon^{\alpha\omega\gamma}\epsilon^{\beta\lambda\nu}+\epsilon^{\alpha\nu\gamma}\epsilon^{\beta\lambda\omega}\right)\chi_{\gamma\lambda},
\end{split}
\end{equation}
where $\epsilon$'s are the Levi-Civita symbols.  

The CRG contribution represents a physical response of the bulk
material to the rotations resulting from such nonlongitudinal strain
gradients. However, in the context of calculating FxE coefficients,
it is useful to remove this contribution. The reasoning for doing this is
based on the fact that, as shown in Ref.~\onlinecite{StengelUNPUB},
the diamagnetic circulating currents from the CRG contribution are
divergentless, and therefore do not result in a build up of charge density
anywhere in the crystal. Therefore, for the experimentally relevant case
of a  \emph{finite} crystal, where the polarization response is completely
determined by the induced charge density, the CRG contribution will not
produce an electrically measurable response.

The fact that the CRG does contribute to the bulk FxE coefficients,
but not to the measurable response of a finite sample, highlights the
fact that, for flexoelectricity, the bulk and surface response are
intertwined\cite{Stengel2014,StengelUNPUB,StengelChapter}. Indeed,
it was determined in Ref.~\onlinecite{StengelUNPUB} that there is
a surface CRG contribution that will exactly cancel the bulk one
[Eq.~(\ref{pchi})]. Thus removing the CRG contribution from the bulk
coefficients simply corresponds to a different way of partitioning the
response between the bulk and the surface. In this work we are focused
on the bulk response, and are free to choose a convention for this
partition. In order to make a more direct connection with experiments,
and to be able to directly compare with charge-density-based calculations
\cite{Stengel2014}, we choose to remove the CRG contribution from our
calculated $\overline{P}^{(2,\omega\nu)}_{\alpha,\kappa\beta}$.

To calculate $\chi_{\gamma\lambda}$, there is again a subtlety
involved in the use of nonlocal pseudopotentials.  Conventional
calculations of the diamagnetic susceptibility involve applying a
vector potential perturbation and calculating the current response
\cite{ICL2001,Pickard2001,Pickard2003,Vignale1991,Mauri1996}.  In the
case of a local Hamiltonian the aforementioned rotational field is
indistinguishable from an electromagnetic vector potential, and the
expression for $\chi_{\gamma\lambda}$ is identical to the diamagnetic
susceptibility. However, in the case of a nonlocal Hamiltonian this is
no longer true.  In that case, the perturbation remains the
\emph{local} current operator, $\hat{\mathcal{J}}^{\text{loc}}$, while
the current response is evaluated using the total (local plus
nonlocal) $\hat{\mathcal{J}}$ (\textit{cf.}~Appendix~\ref{Appdiamag}).
This difference indicates that Larmor's theorem may break
down for nonlocal potentials. This is discussed further in
Sec.~\ref{Disc}.

\section{Implementation\label{Imp}}

The procedure for calculating the FxE coefficients using the formalism
in Sec.~\ref{Form} is as follows. We first perform conventional DFPT
phonon calculations [displacing sublattice $\kappa$ in direction
$\beta$, as in Eq.~(\ref{phon})] at small but finite wavevectors
\textbf{q} to obtain the static first-order wavefunctions
$\vert\partial_{\lambda}
u^{\kappa\beta}_{n\textbf{k},\textbf{q}}\rangle$.  We choose $\vert
q\vert < 0.04$, where here and henceforth we express $q$ in reduced
units of $2\pi/a$ ($a$ is the cubic lattice constant).  To avoid the
sum over empty states in Eq.~(\ref{deltapsi}), we determine the
first-order adiabatic wavefunctions by solving the Sternheimer
equation
\begin{equation}
\label{deltastern}
(H_{\textbf{k}}-\epsilon_{n\textbf{k}})\vert\delta u^{\kappa\beta}_{n\textbf{k},\textbf{q}}\rangle=-i\mathcal{Q}_{c,\textbf{k}+\textbf{q}}\vert\partial_{\lambda} u^{\kappa\beta}_{n\textbf{k},\textbf{q}}\rangle
\end{equation}
where $\epsilon_{n\textbf{k}}$ is the eigenvalue of band $n$ and
$k$-point \textbf{k} and $\mathcal{Q}_{c,\textbf{k}+\textbf{q}}$ is
the projector over conduction band states (implemented as one minus
the projector over valence states).  Then we apply the current
operator in Eq.~(\ref{JqExpand}) to obtain
$\overline{P}_{\alpha,\kappa\beta}^{\textbf{q}}$ from Eq.~(\ref{PqTR})
(see Appendices \ref{Jloc} and \ref{JNL} for details).

As will be discussed in Sec.~\ref{Bench}, we will use the ICL path for
most of the calculations in this study, so the explicit expression for
this case is provided in this section.  The local contribution to
$\overline{P}_{\alpha,\kappa\beta}^{\textbf{q}}$ is derived in
Appendix \ref{Jloc}, leading to Eq.~(\ref{pkq}).
The three terms in the small-${\bf q}$ expansion of the nonlocal part
are determined in Appendix \ref{ICL} by combining Eqs.~(\ref{JkqICL})
and (\ref{PqTR}), and expanding in powers of \textbf{q}, leading to
Eq.~(\ref{Pqexpand2}). Combining Eq.~(\ref{Pqexpand2}) with
  Eqs.~(\ref{ICL0})-(\ref{ICL2}) and adding Eq.~(\ref{pkq}), we have
\begin{widetext}
\begin{equation}
\begin{split}
\label{ICLimp}
\overline{P}_{\alpha,\kappa\beta}^{\textbf{q},\text{ICL}}&=-\frac{4}{N_k}\sum_{n\textbf{k}}\Bigg[\langle u_{n\textbf{k}}\vert\hat{p}_\alpha^{\textbf{k}}+\frac{q_\alpha}{2}\vert\delta u^{\kappa\beta}_{n\textbf{k},\textbf{q}}\rangle+\langle u_{n\textbf{k}}\vert\frac{\partial \hat{V}^{\textbf{k}{,\text{nl}}}}{\partial k_\alpha}\vert\delta u^{\kappa\beta}_{n\textbf{k},\textbf{q}}\rangle
\\
&+\frac{1}{2}\sum_{\gamma=1}^3 q_\gamma\langle u_{n\textbf{k}}\vert\frac{\partial^2 \hat{V}^{\textbf{k},\text{nl}}}{\partial k_\alpha\partial k_\gamma}\vert\delta u^{\kappa\beta}_{n\textbf{k},\textbf{q}}\rangle
+\frac{1}{6}\sum_{\gamma=1}^3\sum_{\xi=1}^3q_\gamma q_\xi\langle u_{n\textbf{k}}\vert\frac{\partial^3 \hat{V}^{\textbf{k},\text{nl}}}{\partial k_\alpha\partial k_\gamma\partial k_\xi}\vert\delta u^{\kappa\beta}_{n\textbf{k},\textbf{q}}\rangle\Bigg],
\end{split}
\end{equation}
\end{widetext}
where we have again assumed TRS [\textit{cf.}~Eq.~\ref{PqTR}].  A
similar equation can be obtained for the PM path using the first-
  and second-order current operators derived in Appendix~\ref{PM}
  [Eqs.~(\ref{PM1}) and (\ref{PM2})].

In order to obtain
$\overline{P}^{(2,\omega\nu)}_{\alpha,\kappa\beta}$, we calculate
numerical second derivatives with respect to $q_\omega$ and $q_\nu$
yielding the needed flexoelectric coefficients
$\mu^{\text{I}}_{\alpha\beta,\omega\nu}$ via Eq.~(\ref{muI}).  Note
that, in addition to the explicit factors of $q$ multiplying the last
two terms, each term has an implicit $q$ dependence through $\delta
u^{\kappa\beta}_{n\textbf{k},\textbf{q}}$
so all terms may contribute to the second derivative.

Since we will consider cubic materials there are three independent FxE coefficients \cite{Hong2013,Stengel2013}:
\begin{equation}
\label{mu}
\begin{split}
&\mu_{\text{L}}=\mu^{\text{II}}_{11,11}=\mu^{\text{I}}_{11,11},
\\
&\mu_{\text{S}}=\mu^{\text{II}}_{12,12}=\mu^{\text{I}}_{11,22},
\\
&\mu_{\text{T}}=\mu^{\text{II}}_{11,22}=2\mu^{\text{I}}_{12,12}-\mu^{\text{I}}_{11,22},
\end{split}
\end{equation}
where L stands for longitudinal, S for shear, and T for transverse.

\tnewpage

\subsection{Electrostatic boundary conditions\label{electroBC}}

The current response to a phonon perturbation, just like other
response properties, displays a strongly nonanalytic behavior in a
vicinity of the $\Gamma$ point (${\bf q}=0$), so some care is
  required when taking the long-wavelength expansions
described in the previous Sections.
A long-wavelength phonon naturally imposes ``mixed'' electrical (ME)
boundary conditions:~\cite{Hong2013} Along the longitudinal direction
($\hat{\bf q}$) the electric displacement field, ${\bf D}$, must
vanish (${\bf D}\cdot \hat{\bf q}=0)$; conversely, periodicity is
preserved in the planes that are normal to $\hat{\bf q}$, resulting in
a vanishing electric field therein.  In general, the bulk FxE tensor
needs to be defined under isotropic ``short-circuit'' (SC) boundary
conditions, which implies that the problematic longitudinal ${\bf
  E}$-fields must be suppressed.  In our calculations, this goal can
be achieved using the procedure of Refs.~\onlinecite{Stengel2013} and
\onlinecite{Stengel2014}, where the $\textbf{G}=0$ component of the
self-consistent first-order potential is removed in the DFPT
calculation of
$\partial_{\lambda}u^{\kappa\beta}_{n\textbf{k},\textbf{q}}$
[Eq.~(\ref{deltastern})]. We will use this procedure for the
calculations of cubic oxides in Sec.~\ref{Cub}.

For several reasons, one may sometimes be interested in calculating
the flexoelectric coefficients under mixed electrical boundary
conditions; in such a case, of course, the $\textbf{G}=0$ component of
the self-consistent first-order potential should not be removed.
Then, however, one must keep in mind that the long-wavelength expansion of
the polarization response is only allowed along a fixed direction in
reciprocal space.
(This implies performing the calculations at points ${\bf q}= q
\hat{\bf q}$, and subsequently operating the Taylor expansion as a
function of the one-dimensional parameter $q$.)
In crystals where the macroscopic dielectric tensor is isotropic and
$\hat{\bf q}$ corresponds to a high-symmetry direction, the
longitudinal coefficients for mixed electrical boundary conditions
are simply related to the short circuit ones by the dielectric
constant, $\epsilon$,
\begin{equation}
\label{BCs}
\mu_{\rm L}^{\rm SC} = \epsilon \mu_{\rm L}^{\rm ME}.
\end{equation}

We will use mixed electrical boundary conditions for our benchmark
calculations of noble gas atoms in Sec.~\ref{Bench} since, in this
particular system, $\mu_{\rm L}^{\rm ME}$, rather than $\mu_{\rm
  L}^{\rm SC}$, can be directly compared to the moments of the
real-space charge density \cite{Hong2013}, as discussed in
Sec.~\ref{IRCmod}.

\subsection{Magnetic susceptibility contribution\label{mag}}
In Sec.~\ref{diamag}, we explained that the diamagnetic susceptibility
is required in order to correct for the CRG contribution to the
FxE coefficients. To avoid the sum over states in Eq.~(\ref{udyn}), we
solve the Sternheimer equation
\begin{equation}
\label{diamagstern}
(\hat{H}_{\textbf{k}}-\epsilon_{n\textbf{k}})\vert\partial_{\dot{\alpha}} u^\alpha_{n\textbf{k},\textbf{q}}\rangle=\mathcal{Q}_{c,\textbf{k}+\textbf{q}}\left(\hat{p}_\alpha^{\textbf{k}}+\frac{q_\alpha}{2}\right)\vert u_{n\textbf{k}}\rangle.
\end{equation}
Recall that
$-\left(\hat{p}_\alpha^{\textbf{k}}+\hat{q}_\alpha/2\right)$ is the
cell-averaged current operator in the case of a local potential.  We
then apply the \emph{full} current operator [Eq.~(\ref{JqExpand})] to
obtain Eq.~(\ref{pdyn}) at several small but finite $q$ (as above,
$\vert q\vert < 0.04$) in order to perform a numerical second
derivative and obtain $\overline{P}^{(2,\omega\nu),\text{
    CRG}}_{\alpha,\beta}$ from Eq.~(\ref{pchi}).

For the case of a material with cubic symmetry, where
$\chi_{\alpha\beta}=\chi_{\text{mag}}\delta_{\alpha\beta}$, we see
from Eq.~(\ref{pchi}) that there will be two nonzero elements of the
CRG contribution: $\overline{P}^{(2,22),\text{
    CRG}}_{1,1}=2\chi_{\text{mag}}$ and $\overline{P}^{(2,12),\text{
    CRG}}_{1,2}=-\chi_{\text{mag}}$.  Therefore, the CI FxE
constants with the CRG contribution removed, $\mu^\prime$, are
given by \cite{StengelUNPUB}
\begin{equation}
\label{mucorr}
\begin{split}
&\mu_{\text{L}}^\prime=\mu_{\text{L}},
\\
&\mu_{\text{S}}^\prime=\mu_{\text{S}}-\chi_{\text{mag}},
\\
&\mu_{\text{T}}^\prime=\mu_{\text{T}}+2\chi_{\text{mag}},
\end{split}
\end{equation}
for cubic materials.

\subsection{Rigid-core correction \label{RCC}}

It was demonstrated in Ref.~\onlinecite{Hong2011} that the CI
FxE constants depend on the treatment of the core density, which will
be different for a different choice of pseudopotential. This
dependence is exactly canceled when the surface contribution is
calculated consistently with the same pseudopotentials
\cite{Stengel2013natcom,StengelChapter}. In order to report more
``portable'' values for the bulk FxE coefficients, we apply the
rigid-core correction (RCC) of Refs.~\onlinecite{Hong2011} and
\onlinecite{Hong2013}:
\begin{equation}
Q^{\text{RCC}}_\kappa=4\pi\int dr r^4 \left[\rho_\kappa^{\text{AE}}(\textbf{r})-\rho_\kappa^{\text{PS}}(\textbf{r})\right],
\end{equation}
where $\rho_\kappa^{\text{AE}}(r)$ is the all-electron density of the
free atom of type $\kappa$, and $\rho_\kappa^{\text{PS}}(r)$ is the
corresponding pseudocharge density. In Table \ref{RCCtab} we list
$Q^{\text{RCC}}$ for the various atoms that we will require for the
cubic oxides reported below (no RCC is included for the noble gas
atoms in Sec.~\ref{Bench}).  Specifically, for short circuit boundary
conditions, $\epsilon\sum_\kappa Q^{\text{RCC}}_{\kappa}/6\Omega$ must
be added to $\mu_{\text{L}}$ and $\mu_{\text{T}}$
\cite{StengelChapter}.

\begin{table}
\caption{$Q^{\text{RCC}}$ for the various atoms in the materials in Sec.~\ref{Cub} in units of e Bohr$^2$.}
\begin{ruledtabular}
\label{RCCtab}
\begin{tabular}{cccc}
& $Q^{\text{RCC}}$ & &$Q^{\text{RCC}}$  \\ \hline
Sr&$-5.93$&Ba &$-13.39$\\
Ti&$-0.54$&Zr &$-4.55$  \\
O &$-0.01$&Pb &$-15.16$\\
Mg &$-4.85$\\

\end{tabular}
\end{ruledtabular}
\end{table}

\subsection{Computational details}
We have implemented the procedure for calculating the FxE coefficients
in the {\sc abinit} code \cite{Abinit_1}. The PBE generalized gradient
approximation functional \cite{pbe} is used throughout. The
conventional phonon and dielectric constant calculations are carried
out using the DFPT implementation available in the code
\cite{Abinit_phonon_1,Gonze1997}. In order to solve the
nonselfconsistent Sternheimer Eqs.~(\ref{diamagstern}) and
(\ref{deltastern}), {\sc abinit}'s implementation of the variational
approach of Ref.~\onlinecite{Gonze1997} is used.

The nuclei and core electrons are described with optimized
norm-conserving Vanderbilt pseudopotentials \cite{Hamann2013} provided
by {\sc abinit}. For the cubic oxides, an $8\times8\times8$
Monkhorst-Pack \cite{Monkhorst1976} $k$-point mesh is used to sample
the Brillouin zone, and the plane-wave energy cutoff is set of 60
Ha. For the isolated atoms, a $2\times2\times2$ $k$-point mesh is used,
and the plane-wave energy cutoff is set of 70 Ha.

\section{Results\label{Res}}

\subsection{Benchmark test: Isolated noble gas atoms\label{Bench}}

\subsubsection{Isolated rigid charge model\label{IRCmod}}

In order to test the implementation described in Sec.~\ref{Imp}, we
consider the toy model of a material made of rigid noninteracting
spherical charge distributions arranged in a simple cubic lattice, as
explored in Refs.~\onlinecite{StengelUNPUB},
\onlinecite{Stengel2013natcom}, and \onlinecite{StengelChapter}.  We
shall refer to this henceforth as the ``isolated rigid charge" (IRC)
model.  Of course, such a material is fictitious, since it would have
no interatomic forces to hold it together; even so, it serves as an
interesting test case since its FxE properties can be determined
analytically and compared to our numerical calculations. In this
section, we will briefly summarize the expectations of the IRC model
(see Refs.~\onlinecite{StengelUNPUB} and
\onlinecite{Stengel2013natcom} for a more complete discussion).

For the IRC ``material,'' there is only one sublattice per cell.  Each
``atom'' is represented by a spherically symmetric charge density
$\rho_{\text{IRC}}(r)$ that falls to zero beyond a cutoff $r_c$ chosen
small enough to ensure that the atomic spheres do not overlap.  The
atoms are assumed to be neutral,
$\int_0^{r_c}\rho_{\text{IRC}}(r)\,r^2\,dr=0$. It was shown in
Ref.~\onlinecite{StengelUNPUB} that the longitudinal and shear
coefficients for the IRC model calculated from the induced current-density
 are
\begin{equation}
\label{muIRC}
 \mu_{\text{L,IRC}}=\mu_{\text{S,IRC}}=\frac{Q_{\text{IRC}}}{2\Omega},
\end{equation}
where $\Omega=a^3$ is the cell volume, and
\begin{equation}
\label{QIRC}
Q_{\text{IRC}}=\int d^3r \rho_{\text{IRC}}(r)x^2
\end{equation}
is the quadrupolar moment of the atomic charge density (of course the
direction $x$ is arbitrary since the charge density is spherically
symmetric).

The FxE constants in Eq.~(\ref{muIRC}) include the CRG
contribution to the current discussed in
Sec.~\ref{diamag}\cite{StengelChapter,
  StengelUNPUB,Stengel2014}. Removing this contribution from our bulk
coefficients [see Eq.~(\ref{mucorr})] results in the primed
coefficients for the IRC model\cite{StengelUNPUB}
\begin{equation}
\label{muprimeIRC}
\mu_{\text{L,IRC}}^\prime=\frac{Q_{\text{IRC}}}{2\Omega},\;\;\;\mu_{\text{S,IRC}}^\prime=0,
\end{equation}
where the CRG contribution is given by
\begin{equation}
\label{XIRC}
\chi_{\text{mag,IRC}}=\mu_{\text{S,IRC}}= \frac{Q_{\text{IRC}}}{2\Omega}
\end{equation}
If we assume that Larmor's theorem holds (i.e., that the CRG
contribution is identical to the magnetic susceptibility),
Eq.~(\ref{XIRC}) is just a statement of the Langevin theory of
diamagnetism, which relates the magnetic susceptibility to the
quadrupole moment of a spherical atomic charge (see Sec.~\ref{Disc}).

\subsubsection{Noble gas atoms\label{noble}}

In the following subsections (\ref{RSmoment}, \ref{tstlong},
\ref{tstshear}), we will compare the behavior of this model with the
results of DFT calculations on isolated noble gas atoms. Several
points should be considered when comparing the results of such
calculations to the expectations of the IRC model (relations in
Sec.~\ref{IRCmod}).

Firstly, the noble gas atoms in our DFT calculations are slightly
polarizable, i.e., not perfectly described by rigid charge
densities. For this reason the longitudinal FxE coefficient will
depend on the choice of electrostatic boundary conditions (see
Sec.~\ref{electroBC}). We will use mixed electrical boundary
conditions, where we should find [analogously to Eq.~(\ref{muIRC})]
\begin{equation}
\label{muNG}
\mu_{\text{L,NG}}^{\text{ME}}=\frac{Q_{\text{NG}}}{2\Omega},
\end{equation}
where the subscript ``NG'' indicates a DFT calculation on a noble
gas atom, and $Q_{\text{NG}}$ is the quadropole moment of the
unperturbed, ground-state charge density of the noble gas atom. If we
had used short circuit boundary conditions, there would have been a
factor of $\epsilon$ on the right-hand side of Eq.~(\ref{muNG}). Of
course, in the IRC model, the ``atoms'' are neutral, rigid, and
spherical, so $\epsilon=1$, and, from Eq.~(\ref{BCs}), short circuit
and mixed electric boundary conditions give the same FxE coefficients.

Also, since our noble-gas-atom calculations will use nonlocal
pseudopotentials, the equality of $\mu_{\text{S,NG}}$ and
$Q_{\text{NG}}/2\Omega$ is not guaranteed; in fact, we will see in
Sec.~\ref{tstshear} that they are not equal. This will
be discussed further in Sec.~\ref{Disc} in the context of the
expected symmetry of the charge response.
 Similarly, we will find
that $\chi_{\text{mag}}$ does not equal $Q_{\text{NG}}/2\Omega$
[\textit{cf}. Eq.~(\ref{XIRC})], indicating that Larmor's theorem
breaks down for our form of the current in the presence of nonlocal
pseudopotentials (discussed in Sec.~\ref{Disc}).

Note that, as with the IRC model, we will drop the $\kappa$ subscript
when discussing the noble gas atoms since the ``crystals'' that we are
considering have only a single sublattice. Also, as all calculations
will use mixed electrical boundary condition, we will drop the
explicit ``ME'' labels.

\subsubsection{Computational strategy: Real-space moments of the charge density\label{RSmoment}}

In addition to the relations in Eqs.~(\ref{muIRC}),
(\ref{muprimeIRC}), and (\ref{XIRC}) of Sec.~\ref{IRCmod} and
Eq.~(\ref{muNG}) of Sec.~\ref{noble}, we can perform specific tests of
the components of our implementation by exploiting the correspondence
between two methods of calculating the FxE coefficients: (i) the
long-wavelength expansion in reciprocal space of the polarization
induced by a phonon [i.e., Eq.~(\ref{muI})] that we have described so
far in this work, and (ii) the computation of the real-space moments
of the induced microscopic polarization or charge density from the
displacement of an isolated atom in a crystal
\cite{Stengel2013,Hong2013}.  For the case of the isolated noble gas
atoms, displacing the entire sublattice (i.e., applying a \textbf{q}=0
acoustic phonon perturbation) is equivalent to displacing a single
atom.

It is particularly useful to compare our methodology to the real-space
moments of the induced charge density, since they can be readily
calculated from a conventional, DFPT phonon calculation (with
$\textbf{q}=0$). Specifically, the longitudinal noble-gas response in
direction $\alpha$ is \cite{Stengel2013,Hong2013}
\begin{equation}
\begin{split}
\label{comp}
\mu_{\text{L,NG}}&=-\frac{1}{2}\frac{\partial^2\overline{P}_{\alpha,\alpha}^{\textbf{q},\text{NG}}}{\partial q_\alpha^2}\Bigg\vert_{\textbf{q}=0}
=\frac{1}{6\Omega}\int_{\text{cell}}d^3r\rho^{\text{NG}}_{\alpha\textbf{q}=0}(\textbf{r})r_\alpha^3.
\end{split}
\end{equation}
where $\rho^{\text{NG}}_{\alpha\textbf{q}}(\textbf{r})\equiv \partial
\rho^{\text{NG}}(\textbf{r})/\partial \lambda_{\alpha\textbf{q}}$ is
the first-order induced charge density from a phonon with wavevector
$\textbf{q}$ and noble gas atoms displaced in the $\alpha$
direction. $\overline{P}_{\alpha,\alpha}^{\textbf{q}}$ is calculated
with mixed electrical boundary conditions. As mentioned in
Sec.~\ref{noble}, the right-hand side of Eq.~(\ref{comp}) equals
$Q_{\text{NG}}/2\Omega$.  Recall that, since the charge density is
related to the divergence of the polarization, it only gives the
longitudinal FxE coefficient. Therefore, we can only use an expression
like the one in Eq.~(\ref{comp}) to test our implementation of
$\mu_{\text{L}}$.

In general (i.e., not specific to the case of the isolated noble gas
  atoms), the induced charge density can be split into contributions
from the local and nonlocal parts of the Hamiltonian, as we did for
the polarization in Eq.~(\ref{PqExpand}). Using the continuity
condition, we can write the first-order charge
as 
\begin{equation}
  \label{delchg}
\rho_{\alpha\textbf{q}}(\textbf{G}+\textbf{q})=-i(\textbf{G}+\textbf{q})\cdot\textbf{P}^{\text{loc}}_{\alpha\textbf{q}}(\textbf{G}+\textbf{q})+\rho_{\alpha\textbf{q}}^{\text{nl}}(\textbf{G}+\textbf{q}) .
\end{equation}
Here $\textbf{P}^{\text{loc}}_{\alpha\textbf{q}}$ is the ``local''
part of the induced polarization and
$\rho_{\alpha\textbf{q}}^{\text{nl}}$ is the nonlocal charge
introduced in Sec.~\ref{contsec}.  Using the reciprocal-space version
of Eq.~(\ref{jloc}), the local induced polarization is (assuming TRS)
\begin{equation}
\label{Ploc}
\begin{split}
P^{\text{loc}}_{\alpha,\alpha\textbf{q}}(\textbf{G}+\textbf{q})=-\frac{2}{N_k}\sum_{n\textbf{k}}\langle \psi_{n\textbf{k}}\vert \left\{ e^{-i(\textbf{G}+\textbf{q})\cdot\hat{\textbf{r}}}, \hat{p}_\alpha\right\}\vert\delta \psi^{\alpha}_{n\textbf{k},\textbf{q}}\rangle
\end{split}
\end{equation}
and the nonlocal charge density from Eq.~(\ref{rhoNL}) is given (in
reciprocal space) by 
\begin{equation}
\label{rhonl}
\begin{split}
\rho^{\text{nl}}_{\alpha\textbf{q}}(\textbf{G}+\textbf{q})=-\frac{4i}{N_k}\sum_{n\textbf{k}}\langle \psi_{n\textbf{k}}\vert \left[ e^{-i(\textbf{G}+\textbf{q})\cdot\hat{\textbf{r}}}, \hat{V}^{\text{nl}}\right]\vert\delta \psi^{\alpha}_{n\textbf{k},\textbf{q}}\rangle
\end{split}
\end{equation}
The first-order charge on the left-hand side of Eq.~(\ref{delchg}) can
be obtained from a conventional DFPT phonon calculation, and thus
Eq.~(\ref{delchg}) allows for several tests of our methodology.

A simple test of the nonlocal contribution at $\textbf{q}=0$ is to
compare the dipole moment of the nonlocal charge with
$\overline{P}_{\alpha,\alpha}^{\textbf{q}\text{,nl}(0)}$ [i.e., the
second term in Eq.~(\ref{ICLimp})], which should give the nonlocal
contribution to the Born effective charge
\begin{equation}
\label{Ztst}
Z^*_{\alpha\beta,\text{nl}}=\overline{P}_{\alpha,\beta}^{\textbf{q}=0,\text{nl}}=\int_{\text{cell}}d^3r\rho^{\text{nl}}_{\beta\textbf{q}=0}(\textbf{r})r_\alpha.
\end{equation}
Again, this relation is generally applicable. For cubic symmetry, the
Born effective charge tensor has only one independent element, which
we write as $Z^*\equiv Z^{*\text{NG}}_{\alpha\alpha}$. Of course, for
the case of the noble gas atom ``material,'' there is only one
sublattice, so the sum of the nonlocal contribution with the local part
(including the ionic charge) will vanish due to the acoustic sum rule
(ASR) \cite{Pick1970}.

For the case of the isolated noble gas atoms, we can use
Eqs.~(\ref{comp}) and (\ref{delchg}) to relate the real-space octupole
moment of $\rho^{\text{nl}}_{\alpha\textbf{q}=0}(\textbf{r})$ [Fourier
transform of Eq.~(\ref{rhonl})] averaged over the cell, to the second
\textbf{q} derivative of
$\overline{P}_{\alpha,\alpha}^{\textbf{q}\text{,nl}}$ [see
Eq.~(\ref{Pqexpand2})] evaluated at
$\textbf{q}=0$. Specifically, we should find that
\cite{Hong2013,Stengel2013}
\begin{equation}
\begin{split}
\label{NLcomp}
-\frac{1}{2}\frac{\partial^2\overline{P}_{\alpha,\alpha}^{\textbf{q},\text{nl,NG}}}{\partial q_\alpha^2}\Bigg\vert_{\textbf{q}=0}=\frac{1}{6\Omega}\int_{\text{cell}}d^3r\rho^{\text{nl,NG}}_{\alpha\textbf{q}=0}(\textbf{r})r_\alpha^3,
\end{split}
\end{equation}
and similarly for the local part,
\begin{equation}
\label{loccomp}
-\frac{1}{2}\frac{\partial^2\overline{P}_{\alpha,\alpha}^{\textbf{q},\text{loc,NG}}}{\partial q_\alpha^2}\Bigg\vert_{\textbf{q}=0}=\frac{1}{6\Omega}\int_{\text{cell}}d^3r\left[-\nabla\cdot\textbf{P}^{\text{loc,NG}}_{\alpha\textbf{q}=0}(\textbf{r})\right]r_\alpha^3,
\end{equation}
where we again perform the reciprocal space calculations using mixed electrical
boundary conditions. 

The comparisons in Eqs.~(\ref{NLcomp}) and (\ref{loccomp}) test both
the long-wavelength expansion of the current operator (local and
nonlocal), and the accuracy of the adiabatic first-order wavefunction
at finite \textbf{q}.

\subsubsection{Test of implementation: Longitudinal response\label{tstlong}}

\begin{figure}
\includegraphics[width=\columnwidth]{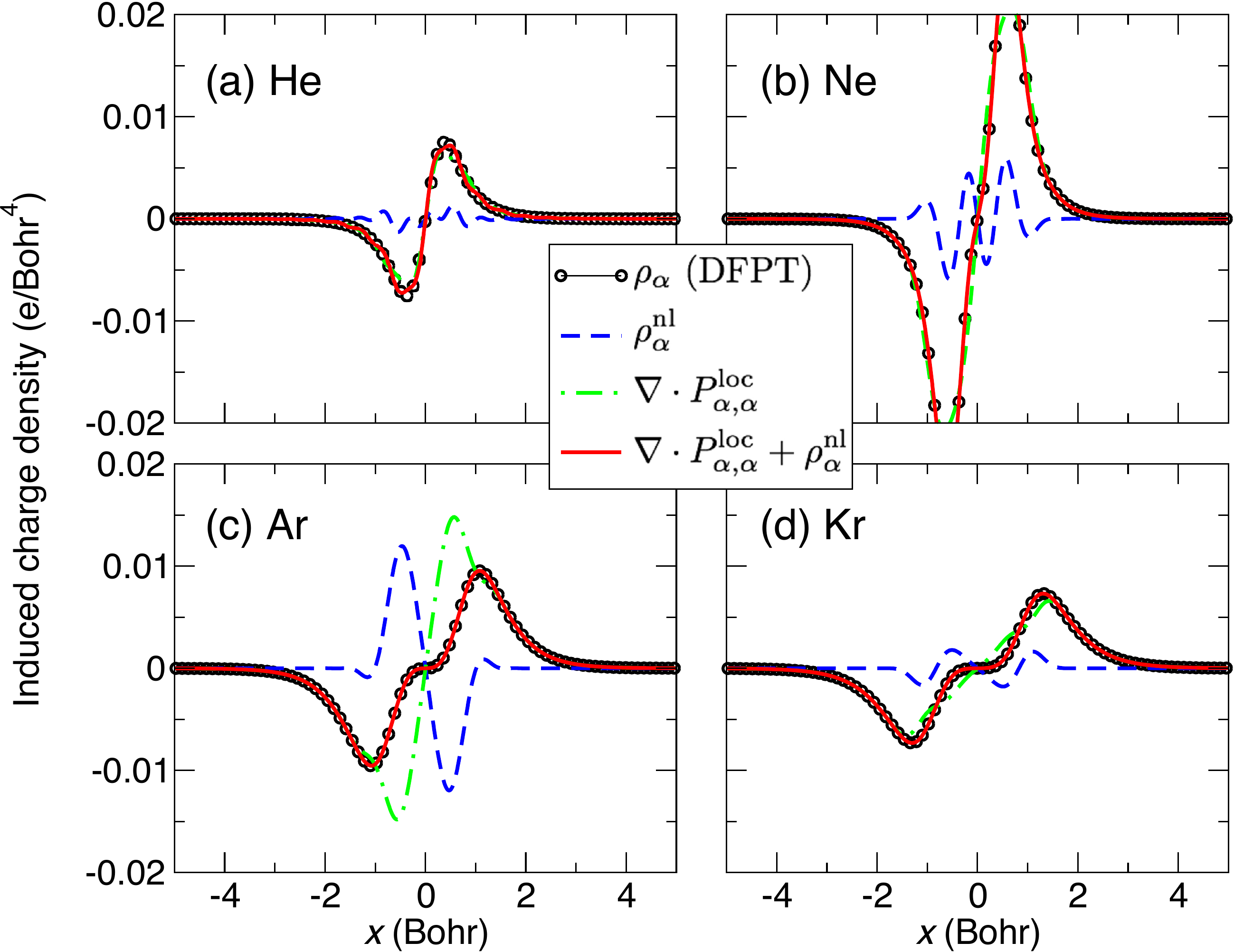}
\caption{\label{NLchg} (Color online) Planar average of the local
  [Eq.~(\ref{Ploc}), green dot-dashed curve], nonlocal
  [Eq.~(\ref{rhonl}), blue dashed], and total [Eq.~(\ref{delchg}), red
  solid] first-order charge for noble gas atoms displaced in the $x$ direction by a $\textbf{q}=0$ phonon. The
  black circles correspond to the first-order charge calculated using
  a conventional, static, DFPT calculation. The box size is
  $16\times16\times16$ Bohr, but zoomed in to only show $\pm 5$
  Bohr.}
\end{figure}

To test $P^{\text{loc}}_{\alpha,\alpha\textbf{q}=0}$ and $\delta
\psi^{\alpha}_{n\textbf{k},\textbf{q}=0}$, we calculate the
first-order charge [left-hand side of Eq.~(\ref{delchg})] from a
$\textbf{q}=0$ phonon by conventional DFPT, and compare to what we
obtain for the right-hand side of Eq.~(\ref{delchg}) calculated using
Eqs.~(\ref{Ploc}) and (\ref{rhonl}) (with $\textbf{q}=0$). We Fourier
transform the quantities in Eq.~(\ref{delchg}) to real space and plot
their planar averages in Fig.~\ref{NLchg} for He, Ne, Ar, and Kr
atoms in $16\times16\times16$ Bohr cells. Summing the contributions
from the nonlocal charge (blue dashed curves) and the gradients of the
local induced polarization (green dot-dashed) gives the red solid
curves in Fig.~\ref{NLchg}. As expected from Eq.~(\ref{delchg}), the
red curve lies on top of the black circles, which correspond to the
first-order charge from the $\textbf{q}=0$ DFPT phonon calculations.

Now we can take the real-space moments of the curves in
Fig.~\ref{NLchg} and compare them with the results of our reciprocal
space expansion.
As discussed in Sec.~\ref{RSmoment}, the first moment of the blue
dashed curves gives the nonlocal contribution to the Born effective
charge, which should correspond to
$\overline{P}_{\alpha,\alpha}^{\textbf{q}=0,\text{nl}}$
[Eq.~(\ref{Ztst})]. In Table \ref{NLchgtab} we give the nonlocal
contribution to $Z^*$ for the noble gas atoms in $14\times14\times14$
Bohr boxes.  The ASR requires that the total $Z^*$ vanishes; for our
noble gas atoms, we calculate the magnitude of the total $Z^*$ to be
less than $10^{-4}$ e, so the ``local'' part (including the
contribution from the ionic charge) is the same magnitude but opposite
sign as the numbers in the second and third columns of Table
\ref{NLchgtab}.

The second column of Table \ref{NLchgtab}, labeled $P^{\text{nl}}$, is
calculated using the reciprocal space current and the third column
(labeled $\rho^{\text{nl}}$) is from the real-space dipole moment of
the charge density. We see that there is excellent agreement between
the two methods, indicating that
$\overline{P}_{\alpha,\alpha}^{\textbf{q}=0,\text{nl}}$ is accurately
calculated.

\begin{table}
  \caption{Calculation of the Born effective charge and $\mu_{\text{L}}$
    using the moments of the local and nonlocal charge (columns labeled
    $\rho$) compared to the current-density implementation (columns
    labeled $P$) for atoms in a $14\times14\times14$ Bohr box. Mixed electrical boundary conditions are used.}
\label{NLchgtab}
\begin{ruledtabular}
\begin{tabular}{c|cc|cccc}
  & \multicolumn{2}{c|}{$Z^*$ (e)} &\multicolumn{4}{c}{$\mu_{\text{L}}$ (pC/m)}\\
& $P^\textrm{nl}$  &$\rho^\textrm{nl}$ &$P^\textrm{loc}$ &$\rho^\textrm{loc}$  &$P^\textrm{nl}$ &$\rho^\textrm{nl}$ \\
\hline	
 He &$-0.027$ &$-0.027$ &$-0.470$ &$-0.470$ &$0.004$  &$0.004$  \\
 Ne &$-0.155$ &$-0.155$ &$-1.872$ &$-1.872$ &$0.028$  &$0.028$ \\
 Ar &$1.556$ &$1.556$ &$-4.620$ &$-4.623$ &$0.073$  &$0.072$ \\
 Kr &$-0.214$ &$-0.214$ &$-5.878$ &$-5.874$ &$-0.099$  &$-0.099$ \\
\end{tabular}
\end{ruledtabular}
\end{table}

It is also clear from Fig.~\ref{NLchg} and Table \ref{NLchgtab}
that the nonlocal correction to the Born effective charge can be very
large, on the order of one electron for Ar. We see a similarly large
contribution for atoms with empty $3d$ shells (but projectors in this
channel) such as a Ca atom or Ti$^{4+}$ ion (not shown).

Now we would like to test the accuracy of our long-wavelength
expansion of the current operator (Sec.~\ref{longwave}) for
calculating $\mu_{\text{L}}$. In Table \ref{NLchgtab} we give both the
local and nonlocal contributions to $\mu_{\text{L}}$ using the
right-hand side of Eqs.~(\ref{NLcomp}) and (\ref{loccomp}) (labeled as
$\rho^{\text{loc}}$ and $\rho^{\text{nl}}$), compared to those
calculated from our current-density implementation [left-hand side of
Eqs.~(\ref{NLcomp}) and (\ref{loccomp}), labeled as $P^{\text{loc}}$
and $P^{\text{nl}}$]. The agreement between the real-space moments and
reciprocal-space derivatives of the expansion in Eq.~(\ref{ICLimp}) is
excellent. Also, we can see that even though the nonlocal contribution
to the Born effective charge is large for Ar, the first-order nonlocal
charge is almost purely dipolar, with the third moment being
almost two orders of magnitude smaller than the contribution of the
local part.

Also, from Table \ref{IRCtab} and Fig.~\ref{IRCplot}, we see that
$\mu_{\text{L}}= Q_{\text{NG}}/2\Omega$ [consistent with
Eq.~(\ref{muNG})] quite accurately for sufficiently large simulation
cells.

\subsubsection{Test of implementation: Shear response\label{tstshear}}

\begin{figure}
\includegraphics[width=\columnwidth]{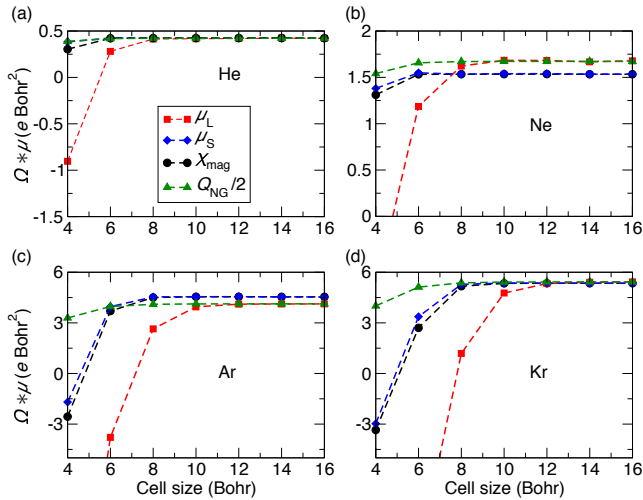}
\caption{\label{IRCplot} (Color online) The longitudinal (red squares)
  and shear (blue diamonds) FxE coefficients, as well as the
  diamagnetic susceptibility correction (black circles) and
  $Q_{\text{NG}}/2\Omega$, for (a) He, (b) Ne, (c) Ar, and (d) Kr
  atoms in cells with various lattice constants. All quantities are
  multiplied by the cell volume, $\Omega$. }
\end{figure}

In Table \ref{IRCtab} we give the longitudinal and shear FxE
coefficients, as well as $\chi_{\text{mag}}$ and
$Q_{\text{NG}}/2\Omega$, for noble gas atoms in $14\times14\times14$
Bohr boxes. For $\mu_{\text{S}}$ and $\chi_{\text{mag}}$, we give
values using the ICL and PM paths for the nonlocal correction.  In
Fig.~\ref{IRCplot}, we show the dependence of these quantities on the
box size.

\begin{table}
  \caption{Longitudinal and shear (ICL and PM path) FxE coefficients for noble gas atoms in $14\times14\times14$ Bohr boxes, as well as the
    diamagnetic susceptibility correction, $\chi_{\text{mag}}$ (ICL and PM path), and the quadrupole
    moment of the unperturbed charge density divided by two times the
    volume [\textit{cf.}~Eqs.~(\ref{muIRC}) and (\ref{QIRC})]. All quantities are in units of pC/m, and mixed electrical boundary conditions used. }
\begin{ruledtabular}
\label{IRCtab}
\begin{tabular}{cccccccc}
&$\mu_{\text{L}}$  &$\mu_{\text{S}}^{\text{ICL}}$&$\mu_{\text{S}}^{\text{PM}}$ & $\chi_{\text{mag}}^{\text{ICL}}$ &$\chi_{\text{mag}}^{\text{PM}}$  &$Q_{\text{NG}}/2\Omega$\\
\hline	
  He &$-0.468$&$-0.467$&$-0.464$ &$-0.468$&$-0.464$&$-0.466$ \\
 Ne &$-1.840$ &$-1.693$ &$-1.655$ &$-1.692$&$-1.655$&$-1.845$ \\
 Ar &$-4.545$  &$-5.008$&$-5.086$ &$-5.013$&$-5.081$&$-4.554$ \\
 Kr &$-5.968$ &$-5.901$ &$-5.917$ &$-5.903$&$-5.921$&$-5.990$ \\ 
\end{tabular}
\end{ruledtabular}
\end{table}

From Table \ref{IRCtab} and Fig.~\ref{IRCplot}, we see that
$\mu_{\text{S}}=\chi_{\text{mag}}$ (consistent with the isotropic
symmetry of the atoms) for sufficiently large simulation
cells.  However, for atoms other than He, $\chi_{\text{mag}}$ is
noticeably different from $Q_{\text{NG}}/2\Omega$, even for large box
sizes. This discrepancy demonstrates that either Larmor's theorem or
the Langevin theory of diamagnetism breaks down when nonlocal
pseudopotentials are present (see Sec.~\ref{Disc} for further
discussion).

When we compare the two path choices, PM (Sec.~\ref{formPM}) and ICL
(Sec.~\ref{formICL}), we find slight quantitative differences for the
shear component and diamagnetic correction. However, the differences
between the paths vanishes for $\mu_{\text{S}}^\prime$ [see
Eq.~(\ref{mucorr})], indicating that although the CRG contribution
is path-dependent, the ``true'' shear response (which is vanishing for
spherical symmetry) is not for this system.  This result is an
excellent test that our implementation is sound. Indeed, for a cubic
solid, all three components of the electronic flexoelectric tensor
$\bm{\mu}'$ can be related to the surface charge accumulated via the
mechanical deformation of a finite crystallite; thus, they should not
depend on the aforementioned path choice.
As the path choice is irrelevant in our context, in the next Section
we shall perform our calculations on cubic oxides using the ICL path.
In Sec.~\ref{Disc} we shall provide a critical discussion of the ICL
and PM prescriptions from a more general perspective, and leave a
detailed comparison of the two approaches for a future work.

\subsection{Cubic oxides\label{Cub}}
We now apply our methodology to calculate the bulk, CI FxE
coefficients for several technologically important cubic oxides. As
mentioned before, we will be using short circuit boundary conditions
and the ICL path for the nonlocal contribution.

\begin{figure}
\includegraphics[width=\columnwidth]{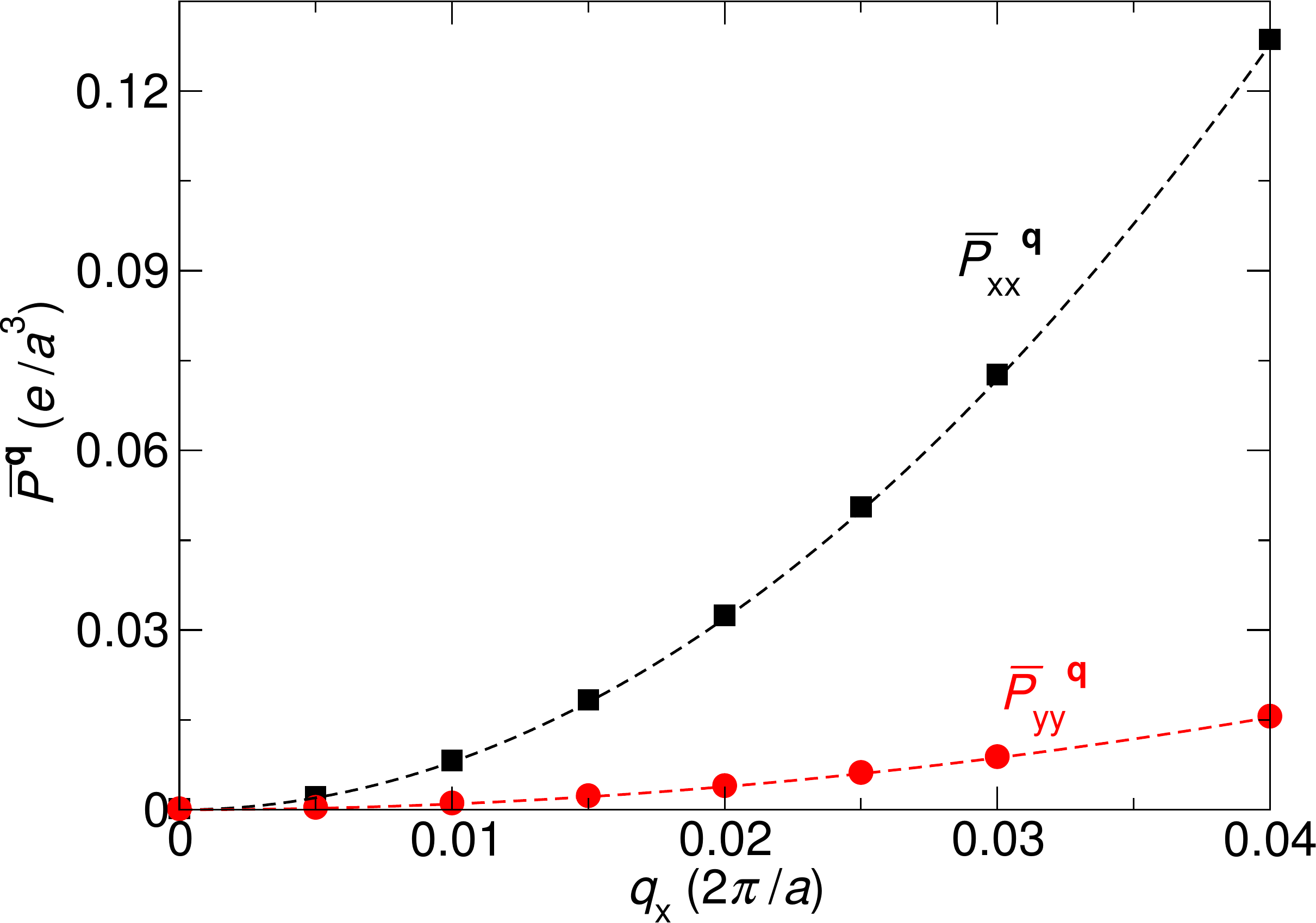}
\caption{\label{STOPq}  (Color online) Induced polarization vs
  $\textbf{q}=(q_x,0,0)$ for cubic SrTiO$_3$. The black (red) points
  correspond to the $x$ ($y$) component of the polarization for atomic
  displacements of the atoms in the $x$ ($y$) direction. Dashed curves
  are quadratic fits. Units are with respect to the calculated SrTiO$_3$ lattice constant $a=7.435$ Bohr.}
\end{figure}

As an example of a typical calculation, in Fig.~\ref{STOPq} we plot
the induced polarization [Eq.~(\ref{ICLimp})] versus
$\textbf{q}=(q_x,0,0)$ for cubic SrTiO$_3$, both for polarization
direction and atomic displacement $\alpha=\beta=x$ and
$\alpha=\beta=y$. As expected, the dependence on $q$ is quadratic
(there is no linear term since cubic SrTiO$_3$ is not piezoelectric
\cite{Hong2013,Stengel2013}), and $\overline{P}^{\textbf{q}}=0$ at
$\textbf{q}=0$, which is required by the ASR condition that the sum of
the Born effective charges should vanish\cite{Pick1970}. By taking the
second derivative of the black (red) dashed curves in
Fig.~\ref{STOPq}, we can obtain $\mu_{11,11}^{\text{I}}$
($\mu_{11,22}^{\text{I}}$). The remaining coefficient
$\mu_{12,12}^{\text{I}}$ is obtained by calculating
$\overline{P}^{\textbf{q}}_{12}$ at various $\textbf{q}=(q_x,q_y,0)$,
and performing a numerical mixed derivative $\partial^2/\partial
q_x\partial q_y$ (not shown).

In Table \ref{cubtab}, we give the FxE coefficients corrected for the
CRG contribution [\textit{cf.} Eq.~(\ref{mucorr})] and the RCC
(Sec.~\ref{RCC}). As discussed above, the RCC is added to the
longitudinal and transverse coefficients \cite{StengelChapter}. Note
that the reported $\chi_{\text{mag}}$ is given in pC/m, whereas other
quantities are in nC/m, so this correction is quite small for the
materials calculated. The contribution of the nonlocal potentials to
the FxE coefficients in Table \ref{cubtab}, which are computed using
the ICL path of Appendix \ref{ICL}, represents a more significant
correction than was the case in Sec.~\ref{Bench}: they are in the
range of $0.03$ to $0.12$ nC/m for the longitudinal and transverse
coefficients, and in the range of $-0.02$ to $0.008$ nC/m for the
shear coefficients.
  
\begin{table*}
  \caption{ Lattice constant, CI dielectric constant, rigid-core correction, and longitudinal, transverse, and shear CI FxE
    coefficients (under short circuit boundary conditions), as well as the diamagnetic susceptibility in units
    of nC/m. The FxE constants include the CRG correction (Sec.~\ref{diamag}) and RCC (Sec.~\ref{RCC}).}
\begin{ruledtabular}
\label{cubtab}
\begin{tabular}{cccccccc}
     & $a$ (Bohr) &$\epsilon$&RCC&$\mu^\prime_{\text{L}}$ &$\mu^\prime_{\text{T}}$ &$\mu^\prime_{\text{S}}$&$\chi_{\text{mag}}\times 10^{3}$ \\
\hline
SrTiO$_3$ &7.435&6.191 &$-0.049$&$-0.87$ ($-0.9$\footnotemark[1],$-0.88$\footnotemark[2]) & $-0.84$ ($-0.83$\footnotemark[2]) &$-0.08$($-0.08$\footnotemark[2])&$-7.3$ \\
BaTiO$_3$ &7.601& 6.657 &$-0.107$& $-1.01$ ($-1.1$\footnotemark[1])&$-0.99$ &$-0.08$& $-1.7$ \\
SrZrO$_3$ &7.882 &4.558 & $-0.049$& $-0.63$ &$-0.58$ & $-0.05$&$-36.0$ \\
PbTiO$_3$ &7.496 & 8.370&$-0.158$ &$-1.39$ ($-1.5$\footnotemark[1]) &$-1.35$ &$-0.09$ & $-22.4$\\
 MgO   &8.058 &3.148&$-0.015$ &$-0.28$ ($-0.3$\footnotemark[1]) & $-0.30$ &$-0.07$ & $-66.1$\\
\end{tabular}
\end{ruledtabular}
\footnotetext[1]{Reference~\onlinecite{Hong2013}}
\footnotetext[2]{Reference~\onlinecite{Stengel2014}}
\end{table*}

The only material for which first-principles calculations of the
transverse and shear coefficients are available (in parentheses in
Table \ref{cubtab}) is SrTiO$_3$, and our values are in excellent
agreement with those previous calculations \cite{Stengel2014}.

For all of the materials, the longitudinal and transverse responses
are of similar magnitude, and the shear response is significantly
smaller. This is a similar trend to that of the isolated noble gas
atoms and of the IRC model [\textit{cf.}~Eq.~(\ref{muprimeIRC})],
suggesting that the response is dominated by the ``spherical''
contribution.  The behavior of the cubic oxides differ significantly
from the IRC model, however, when it comes to the contribution of the
CRG correction $\chi_{\text{mag}}$.  For isolated atoms,
$\chi_{\text{mag}}$ is equal to $\mu_{\text{IRC,S}}$, and is of the
same order as $\mu^\prime_{\text{IRC,L}}$; therefore, a vanishing
value of $\mu^\prime_{\text{IRC,S}}$ is only obtained after removing
the CRG contribution [Eq.~(\ref{mucorr})]. In the case of the
cubic oxides, the CRG correction is only a minor contribution to
$\mu^\prime_{\text{S}}$, and $\chi_{\text{mag}}$ is two orders of
magnitude smaller than $\mu^\prime_{\text{L}}$.  In fact,
$\chi_{\text{mag}}$ for the cubic oxides is comparable to that of the
isolated atoms, while the FxE coefficients for the cubic oxides are
two orders of magnitude larger. This indicates that although the
bonding of atoms in the cubic compounds significantly enhances the FxE
coefficients, it does not have a large effect on the CRG
correction.

It should be noted that the value of $\chi_{\text{mag}}$ for SrTiO$_3$
($-2.28\times10^{-7}$ cm$^3$/g after unit conversion) is in fair
agreement with the measured diamagnetic susceptibility of around
$-1\times10^{-7}$ cm$^3$/g from Ref.~\onlinecite{Frederikse1966}.

\section{Discussion\label{Disc}}

Before closing, it is useful to recap the technical issues
that are associated with the calculation of the current density
response in a nonlocal pseudopotential context, and critically
discuss them in light of the result presented in this work.
In particular, it is important to clarify whether our proposed 
approach matches the expectations, especially regarding the
known transformation properties of the current density upon
rototranslations, or whether there is any deviation that needs to
be kept in mind when computing flexoelectric coefficients and
other current-related linear-response properties.

As we have already discussed at length in the earlier Sections,
our definition of the current density (i) satisfies the continuity
equation by construction, (ii) correctly reduces to the textbook formula
in the region of space where the Hamiltonian is local, and (iii) is consistent
with the known formula for the macroscopic current operator.
However, we have not yet discussed some additional properties of the
current density that were established in earlier works, that might be
used as ``sanity checks'' of our implementation:
\begin{itemize}
\item Translational invariance of the charge-density response:
    As established by Martin \cite{Martin1972}, simultaneous uniform 
translation of all atoms in the crystal must yield the same variation in charge
    density at every point as if the static charge density were
    rigidly shifted. Therefore, if the whole crystal undergoes
  a translation with uniform velocity ${\bf v}$, the
  current density in the laboratory frame must be
\begin{equation}
{\bf J}({\bf r}) = {\bf v} \rho({\bf r}),
\label{transl}
\end{equation}
where $\rho({\bf r})$ is the static charge density.
\item Larmor's theorem: The circulating currents generated in a
  crystallite by a uniform rotation with constant angular velocity
  $\bm{\omega}$ (as observed in the frame of the rotating
    material) are, in the linear limit of small velocities, identical
  to the orbital currents that would be generated by an applied (and
  constant in time) ${\bf B}$-field. As a corollary, the rotational
  $g$-factor of closed-shell molecules corresponds to their
  paramagnetic susceptibility.
\item Langevin's diamagnetism: The magnetic susceptibility of a
  spherically symmetric atom is proportional to the quadrupolar moment
  of its ground-state charge density.
\end{itemize}
In the following, we shall analyze how our formalism stands in
relationship to these latter ``weak'' [compared to the ``strong''
conditions (i-iii) above] criteria of validity.
(By ``weak'' we mean not required for a
physically sound calculation of the flexoelectric tensor, but possibly
 necessary for a wider range of physical properties.)

\subsection{Translational invariance of the charge-density response}

Based on our results of Table III, we can safely conclude that both
flavors of the current-density operator (ICL and PM) break translational 
invariance, Eq.~(\ref{transl}). 
To see this, consider the shear flexoelectric coefficient of an 
isolated atom in a box, (e.g., $\mu_{\rm S,NG}$). This quantity can be  
defined in real space as the second moment of the microscopic
current-density response to the displacement of an isolated atom,
\begin{equation}
\mu_{\rm S} = \frac{1}{2\Omega} \int d^3 r \frac{\partial J_y({\bf r})}{\partial \dot{\lambda}_y} x^2,
\label{mus-jr} 
\end{equation}
where $\dot{\lambda}_y$ stands for the velocity of the atom along $y$.
This formula, as it stands, is not very practical for calculations:
our implementation does not allow for a fully microscopic 
calculation of ${\bf J}({\bf r})$, and therefore we had to 
replace Eq.~(\ref{mus-jr}) with computationally more tractable
small-${\bf q}$ expansions. 
Still, Eq.~(\ref{mus-jr}) is quite useful for our purposes, as it
allows us to draw general conclusions about ${\bf J}({\bf r})$ without
the need for calculating it explicitly.
In particular, if translational invariance [Eq.~(\ref{transl})] were
satisfied, then we could plug Eq.~(\ref{transl}) into
  Eq.~(\ref{mus-jr}) and use Eq.~(\ref{QIRC}) to obtain $\mu_{\rm S,NG}
  = \frac{1}{2\Omega}\int d^3r \rho(\textbf{r}) x^2
  =Q_{\text{NG}}/2\Omega$.  [This equality is a necessary but not
sufficient condition for the validity of Eq.~(\ref{transl}).]
As we can see from Table III, $\mu_{\rm S,NG}$ is only approximately
equal to $Q_{\text{NG}}/2\Omega$ for both the ICL and PM flavors of
the current-density operator.
This implies that neither approach is able to guarantee translational
invariance. 

Similarly, the data we have in hand
does not allow us to establish a clear preference between the PM and
ICL recipes, as the discrepancies between the two are typically much
smaller (and devoid of a systematic trend) than their respective
failure at satisfying $\mu_{\rm S,NG} = Q_{\text{NG}}/2\Omega$.  Note that the discrepancy
strictly consists of \emph{solenoidal} (i.e., divergenceless)
contributions to the current response; the longitudinal components are
exactly treated, as one can verify from the excellent match between
the longitudinal coefficient, $\mu_{\text{L}}$, and the quadrupolar
estimate in Table III.

\subsection{Langevin diamagetism and Larmor's theorem}
We come now to the assessment of the Larmor and Langevin results.
One of the virtues of the PM recipe resides in its superior accuracy
when comparing the orbital magnetic response to all-electron data.
Indeed, in the context of our discussion, one can verify that it 
exactly complies with Langevin's theory of diamagnetism in 
the case of isolated spherical atoms.
\footnote{This can be deduced from Eq.(12) and (13) of Ref.~\onlinecite{Pickard2003}:
By placing a single spherical pseudoatom in the gauge origin, all nonlocal
contributions vanish by construction as they are multiplied by ${\bf R}$;
thus, the applied magnetic field enters the Hamiltonian via the usual
substitution ${\bf p} \rightarrow {\bf p} + {\bf A}$. Then, the first order 
Hamiltonian is the angular momentum operator, which commutes with the
ground-state density matrix and yields a vanishing linear response, and 
the second-order piece picks the quadrupolar moment of the ground-state density,
as in the local case.}
The situation, however, is not so bright regarding Larmor's theorem.
If the latter were satisfied, then the ``rotational orbital susceptibility'' 
$\chi_{\rm mag}$ would match Langevin's quadrupolar expression, as we know that 
Langevin's result holds in the case of a ``true'' ${\bf B}$-field. 
By looking, again, at Table III, we clearly see that this is not the case --
again, there is a discrepancy between the last column (based on the static 
quadrupole) and the calculated values of $\chi_{\rm mag}$.
Since the deviations in $\chi_{\rm mag}$ and $\mu_{\rm S}$ are essentially
identical in the limit of an isolated atom in a box, it is reasonable to 
assume that the underlying factors are similar.

It should be noted that our value for Ne (after unit conversion, ICL path) is
$\chi_{\text{mag}}^{\text{ICL}}=-7.29\times10^{-6}$ cm$^3$/mole, which is fairly
close in magnitude to previously calculated values of the diamagnetic
susceptibility of Ne: $-7.75\times10^{-6}$ cm$^3$/mole\cite{ICL2001}
and $-7.79\times10^{-6}$ cm$^3$/mole\cite{Mauri1996}.

\subsection{Unphysical spatial transfer resulting from nonlocal pseudopotentials}

The reason why the current density violates both translational
invariance and Larmor's theorem has to be sought in the unphysical
transfer of density that can result from the presence of a
  nonlocal potential. That is, a nonlocal operator may project the
  wavefunction (and therefore the particle amplitude) from a point ${\bf
    r}$ to a distant point ${\bf r}'$ in a discontinuous manner, such
  that no current flows through a given surface surrounding ${\bf
    r}$ even though the charge density within that surface changes. Of course,
  this is just a conceptual way of describing the violation of the
  continuity equation, discussed in Sec.~\ref{curden}.

Taking the example of a single atom placed at $\textbf{R}=0$ and using
the PM approach, it is shown in Appendix \ref{appDiv} that the current
density can be written as
\begin{equation}
{\bf J}^{\rm nl}({\bf r}) \sim \frac{\hat{\bf r} C(\hat{\bf r})}{r^2} .
\end{equation}
where $C(\hat{\textbf{r}})$ is a direction-dependent constant that
depends on the nonlocal charge [Eq.~(\ref{Cr})]. Therefore, the
current-density field diverges near the atomic site,
$\textbf{r}\rightarrow0$, and such a divergence can have a different
prefactor and sign depending on the direction.

A diverging ${\bf J}$-field is problematic to deal with and
unphysical. One can easily realize that this characteristic is
incompatible, for example, with the correct transformation laws of
${\bf J}$ under rigid translations.
In particular, the electronic charge density is always finite in a
vicinity of the nucleus, even in the all-electron case where the
corresponding potential does, in fact, diverge.
This implies that Eq.~(\ref{transl}) cannot be satisfied by a
diverging ${\bf J}$-field.

For the ICL path, the nonlocal current does not have such a simple
relation to the nonlocal charge as in the case of the PM path [Eq.~(E4)];
therefore a similar derivation as in Appendix E may not be possible
for the ICL case. However, our numerical results in Table III are sufficient to
conclude that the ICL path violates translational symmetry as well. The
extent of the violation can be quantified by looking at the discrepancy
between $\mu_{\text{L}}$ and $\mu_{\text{S}}$, which is comparably
large in the PM and ICL cases---recall that these two values should,
in principle, coincide for the isolated spherical atoms model.

At present it is difficult to predict whether it might be possible to
cure the above drawbacks by simply choosing a different path for the
definition of the current operator, or whether these difficulties may
require a deeper revision of the nonlocal pseudopotential theory in
contexts where the microscopic current density is needed.
In any case, the flexoelectric coefficients we calculated in this work
for cubic materials are unaffected by these issues: Once the
``diamagnetic'' contribution has been removed, the three independent
coefficients are all well defined in terms of the charge-density
response.
Nonetheless, the above caveats should be kept in mind when using the
present current-density implementation to access flexoelectric
coefficients in less symmetric materials, or other response properties
that depend on the microscopic current response.

\section{Conclusions\label{Con}}

We have developed a DFPT implementation for calculating the bulk
CI flexoelectric tensor from a single unit cell. Therefore,
we have overcome the limitations of previous implementations
(Refs.~\onlinecite{Hong2013} and \onlinecite{Stengel2014}), which
required supercells to calculate the transverse and shear CI
FxE coefficients.

Our implementation is based on calculating the microscopic current
density resulting from the adiabatic atomic displacements of a
long-wavelength acoustic phonon. We have determined a form for the
current-density operator that satisfies the continuity condition in
the presence of nonlocal, norm-conserving pseudopotentials, and
reduces to the correct form in the limit of a uniform, macroscopic
perturbation, and/or when only local potentials are present.

In order to benchmark our methodology, we have used noble gas atoms to
model systems of noninteracting spherical charge densities. The tests
demonstrate the accuracy of our nonlocal correction to the current
operator, as well as the calculated  CRG corrections derived in
Ref.~\onlinecite{StengelUNPUB}. For our form of the current density,
we demonstrate that nonlocal pseudopotentials result in a violation of
translational invariance and Larmor's theorem, though this does not
affect our FxE coefficients after the CRG contribution has been
removed. Finally, we have applied our methodology to several cubic
oxides, all of which show similar trends in that the longitudinal and
transverse responses are similar ($\sim1$ nC/m), and the shear
response is an order of magnitude smaller.

Combining the methodology of this paper with DFPT implementations for
calculating the lattice-mediated contribution to the bulk FxE
coefficients \cite{Stengel2013,Stengel2014}, and the surface
contribution \cite{Stengel2014}, will allow for efficient calculation
of the full FxE response for a variety of materials.

\begin{acknowledgements}
  We are grateful to K.~M.~Rabe, D.~R.~Hamann, B.~Monserrat,
  H.~S.~Kim, A.~A.~Schiaffino, C.~J.~Pickard, and S.~Y.~Park for useful
  discussions. CED and DV were supported by ONR Grant
  N00014-16-1-2951.  MS acknowledges funding from from the European
  Research Council (ERC) under the European Union's Horizon 2020
  research and innovation programme (Grant Agreement No. 724529), and
  from Ministerio de Econom\'{i}a, Industria y Competitividad
  (MINECO-Spain) through Grants No. MAT2016-77100-C2-2-P and
  SEV-2015-0496, and from Generalitat de Catalunya through Grant
  No. 2017 SGR1506.

\end{acknowledgements}

\clearpage

\begin{widetext}

\appendix

\section{Essin \textit{et al.} approach and the Ismail-Beigi, Chang, and Louie straight-line path\label{sepICL}}

Here we perform a long-wavelength expansion of the current operator
using the approach of Essin \textit{et al.}\cite{Essin2010}, and
confirm that the approach is equivalent to that of ICL
\cite{ICL2001}. We start from Eq.~(\ref{JqSL}) and rewrite it as
\begin{equation}
\label{LocNL1}
\begin{split}
\langle\textbf{s}\vert\hat{\mathcal{J}}_{\alpha}^{\text{ICL}}(\textbf{q})\vert\textbf{s}^\prime\rangle&=-iH(\textbf{s},\textbf{s}^\prime)(s_\alpha-s_\alpha^\prime)\frac{e^{-i\textbf{q}\cdot\textbf{s}}-e^{-i\textbf{q}\cdot\textbf{s}^{\prime}}}{i\textbf{q}\cdot(\textbf{s}-\textbf{s}^\prime)}
\\
&=-i\left[\hat{r}_\alpha,\hat{H}\right]_{\textbf{s}\textbf{s}^\prime}\frac{e^{-i\textbf{q}\cdot\textbf{s}}-e^{-i\textbf{q}\cdot\textbf{s}^{\prime}}}{i\textbf{q}\cdot(\textbf{s}-\textbf{s}^\prime)}
\\
&=-\left(i\left[\hat{r}_\alpha,\hat{T}\right]_{\textbf{s}\textbf{s}^\prime}+i\left[\hat{r}_\alpha,\hat{V}^{\text{nl}}\right]_{\textbf{s}\textbf{s}^\prime}\right)\frac{e^{-i\textbf{q}\cdot\textbf{s}}-e^{-i\textbf{q}\cdot\textbf{s}^{\prime}}}{i\textbf{q}\cdot(\textbf{s}-\textbf{s}^\prime)},
\end{split}
\end{equation}
where $\hat{T}$ is the kinetic energy operator and
$\hat{V}^{\text{nl}}$ is the nonlocal part of the potential (the local
part of the potential does not contribute). We now factor out a
$e^{-i\textbf{q}\cdot\textbf{s}^\prime}$ and then expand the term
outside of the parentheses
\begin{equation}
\label{LocNL2}
\begin{split}
\langle\textbf{s}\vert\hat{\mathcal{J}}_{\alpha}^{\text{ICL}}(\textbf{q})\vert\textbf{s}^\prime\rangle&=-\left(i\left[\hat{r}_\alpha,\hat{T}\right]_{\textbf{s}\textbf{s}^\prime}+i\left[\hat{r}_\alpha,\hat{V}^{\text{nl}}\right]_{\textbf{s}\textbf{s}^\prime}\right)e^{-i\textbf{q}\cdot\textbf{s}^\prime}\left(-1+\frac{i\textbf{q}\cdot(\textbf{s}-\textbf{s}^\prime)}{2}+\frac{[\textbf{q}\cdot(\textbf{s}-\textbf{s}^\prime)]^2}{6}+...\right).
\end{split}
\end{equation}
As mentioned in Sec.~\ref{formICL}, if $\textbf{q}=0$, then $\hat{\mathcal{J}}_{\alpha}^{\text{ICL}}(\textbf{q}=0)=i\left[\hat{r}_\alpha,\hat{H}\right]=-\hat{v}_\alpha$, the velocity operator.

Consider the case of a Hamiltonian with a local potential, so the only
term in Eq.~(\ref{LocNL2}) is the commutator of the position operator
with the kinetic part of the Hamiltonian. We can rewrite this term as
\begin{equation}
\label{LocNL3}
\begin{split}
\langle\textbf{s}\vert\hat{\mathcal{J}}_{\alpha}^{\text{loc}}(\textbf{q})\vert\textbf{s}^\prime\rangle&=-\left(-i\left[\hat{r}_\alpha,\hat{T}\right]_{\textbf{s}\textbf{s}^\prime}-\sum_{\gamma=1}^3 \frac{q_\gamma}{2}\left[\hat{r}_\gamma,\left[\hat{r}_\alpha,\hat{T}\right]\right]_{\textbf{s}\textbf{s}^\prime}+\sum_{\gamma=1}^3\sum_{\xi=1}^3 \frac{iq_\gamma q_\xi}{6}\left[\hat{r}_\xi,\left[\hat{r}_\gamma,\left[\hat{r}_\alpha,\hat{T}\right]\right]\right]_{\textbf{s}\textbf{s}^\prime}+...\right)e^{-i\textbf{q}\cdot\textbf{s}^\prime}.
\end{split}
\end{equation}
The term at zeroth order in $q$ is simply the momentum operator:
$\hat{p}_\alpha=-i\left[\hat{r}_\alpha,\hat{T}\right]$; at first order
in $q$, we have $\hat{q}_\alpha/2$ (the nested commutator is simply
the Kroneker delta function $-\delta_{\alpha\gamma}$); higher order
terms vanish.  So in the case of a Hamiltonian that only has a local
potential,
\begin{equation}
\label{LocNL4}
\begin{split}
\hat{\mathcal{J}}_{\alpha}^{\textbf{q},\text{loc}}=-\left(\hat{p}_\alpha+\frac{q_\alpha}{2}\right),
\end{split}
\end{equation}
which is the cell-periodic momentum operator for the case of local
potentials, as we derive in Appendix \ref{Jloc}. Therefore the local
and nonlocal components can be cleanly separated. The nonlocal part of
the potential in Eq.~\ref{LocNL1} is addressed in Appendix \ref{ICL}.

Note that the approach of Essin \textit{et al.} does not work for an
arbitrary choice of path. Specifically, if we were to use
Eq.~(\ref{HA1}) with the PM path choice
$\textbf{s}^\prime\rightarrow\textbf{R}\rightarrow\textbf{s}$, the
expression would not reproduce the correct form of the current for
local potentials (except for the case of the longitudinal
response). Of course, in the PM form of the coupled Hamiltonian in
Eq.~(\ref{HPM}), the current in the case of only local potentials
trivially reduces to the correct form
$\hat{\mathcal{J}}_\alpha^{\text{loc}}(\textbf{r})=-\frac{1}{2}\left\{\hat{\rho}(\textbf{r}),(\hat{p}_\alpha+\hat{A}_\alpha)\right\}$.

\section{Derivation of induced polarization: Local potentials\label{Jloc}}
In this section we derive
$P^{\textbf{q},\text{loc}}_{\alpha,\kappa\beta}$ for
Eq.~(\ref{PqExpand}). This is a straightforward generalization of what
was derived by Umari, Dal Corso, and Resta \cite{Umari2001} to finite
\textbf{q} perturbations, and has been derived previously in other
contexts (e.g. for determining magnetic\cite{Mauri1996} or
dielectric\cite{Adler1962} susceptibility, and in the context of
phonon deformation potentials\cite{Khan1984}).

Using the adiabatic expansion of the time-dependent wavefunction
[Eqs.~(\ref{psiad}) and (\ref{deltapsi})], to first order in
$\dot{\lambda}$ we can write the density matrix as
\begin{equation}
\begin{split}
\label{rho}
\rho(t)&=-\frac{2}{N_k}\sum_{n\textbf{k}}\vert\Psi_{n\textbf{k}}(\lambda(t))\rangle\langle\Psi_{n\textbf{k}}(\lambda(t))\vert\simeq-\frac{2}{N_k}\sum_{n\textbf{k}}\left[\vert\psi_{n\textbf{k}}\rangle\langle\psi_{n\textbf{k}}\vert+\dot{\lambda}(\vert\delta\psi_{n\textbf{k}}\rangle\langle\psi_{n\textbf{k}}\vert+\vert\psi_{n\textbf{k}}\rangle\langle\delta\psi_{n\textbf{k}}\vert)\right]
\end{split}
\end{equation}
where the factor of two is assuming spin degeneracy. If we apply the
local current-density operator [Eq.~(\ref{jloc})], retaining terms
only to linear order in $\dot{\lambda}$, and take the derivative with
respect to $\dot{\lambda}$ we obtain the induced polarization
\begin{equation}
\begin{split}
P_\alpha^{\text{loc}}(\textbf{r})=-\frac{1}{N_k}\sum_{n\textbf{k}}\big[\langle\psi_{n\textbf{k}}\vert\textbf{r}\rangle\langle\textbf{r}\vert\hat{p}_\alpha\vert\delta\psi_{n\textbf{k}}\rangle
+\langle\delta\psi_{n\textbf{k}}\vert\textbf{r}\rangle\langle\textbf{r}\vert\hat{p}_\alpha\vert\psi_{n\textbf{k}}\rangle+\langle\psi_{n\textbf{k}}\vert\hat{p}_\alpha\vert\textbf{r}\rangle\langle\textbf{r}\vert\delta\psi_{n\textbf{k}}\rangle
+\langle\delta\psi_{n\textbf{k}}\vert\hat{p}_\alpha\vert\textbf{r}\rangle\langle\textbf{r}\vert\psi_{n\textbf{k}}\rangle\big] .
\end{split}
\label{pr}
\end{equation}

Now consider the perturbation in Eq.~(\ref{phon}): the displacement of
a sublattice $\kappa$ in direction $\beta$ modulated by a phase with
wavevector \textbf{q}. We begin with the real-space expression for the
polarization induced by this perturbation:
\begin{equation}
P^{\textbf{q},\text{loc}}_{\alpha,\kappa\beta}(\textbf{r})=-\frac{1}{N_k}\sum_{n\textbf{k}}\left[\langle\psi_{n\textbf{k}}\vert\textbf{r}\rangle\langle\textbf{r}\vert\hat{p}_\alpha\vert\delta\psi_{n\textbf{k},\textbf{q}}^{\kappa\beta}\rangle
+\langle\delta\psi_{n\textbf{k},-\textbf{q}}^{\kappa\beta}\vert\textbf{r}\rangle\langle\textbf{r}\vert\hat{p}_\alpha\vert\psi_{n\textbf{k}}\rangle
+\langle\psi_{n\textbf{k}}\vert\hat{p}_\alpha\vert\textbf{r}\rangle\langle\textbf{r}\vert\delta\psi_{n\textbf{k},\textbf{q}}^{\kappa\beta}\rangle
+\langle\delta\psi_{n\textbf{k},-\textbf{q}}^{\kappa\beta}\vert\hat{p}_\alpha\vert\textbf{r}\rangle\langle\textbf{r}\vert \psi_{n\textbf{k}}\rangle
\right]
\label{pqr}
\end{equation}
where the subscript \textbf{q} in
$\delta\psi_{n\textbf{k},\pm\textbf{q}}^{\kappa\beta}$ indicates that
the perturbation couples states at \textbf{k} to those at
$\textbf{k}\pm\textbf{q}$. If we assume TRS [see Eq.~(\ref{TReq})],
then we have
\begin{equation}
P^{\textbf{q},\text{loc}}_{\alpha,\kappa\beta}(\textbf{r})=-\frac{2}{N_k}\sum_{n\textbf{k}}\left[\langle\psi_{n\textbf{k}}\vert\textbf{r}\rangle\langle\textbf{r}\vert\hat{p}_\alpha\vert\delta\psi_{n\textbf{k},\textbf{q}}^{\kappa\beta}\rangle
+\langle\psi_{n\textbf{k}}\vert\hat{p}_\alpha\vert\textbf{r}\rangle\langle\textbf{r}\vert\delta\psi_{n\textbf{k},\textbf{q}}^{\kappa\beta}\rangle
\right]
\label{pqrTRS}
\end{equation}
We Fourier transform Eq.~(\ref{pqrTRS}) to
reciprocal space and consider the cell periodic part
\begin{equation}
\begin{split}
P^{\text{loc}}_{\alpha,\kappa\beta}(\textbf{G}+\textbf{q})&=-\frac{2}{N_k}\sum_{n\textbf{k}}\int d^3r\Big[\langle\psi_{n\textbf{k}}\vert\textbf{r}\rangle e^{-i(\textbf{G}+\textbf{q})\cdot\textbf{r}} \langle\textbf{r}\vert\hat{p}_\alpha\vert\delta\psi^{\kappa\beta}_{n\textbf{k},\textbf{q}}\rangle
+\langle\psi_{n\textbf{k}}\vert\hat{p}_\alpha\vert\textbf{r}\rangle e^{-i(\textbf{G}+\textbf{q})\cdot\textbf{r}}\langle\textbf{r}\vert\delta\psi^{\kappa\beta}_{n\textbf{k},\textbf{q}}\rangle\Big].
\end{split}
\end{equation}
We now explicitly insert the expansion of the wavefunctions in terms
of plane waves
\begin{equation}
\begin{split}
\psi_{\textbf{k}}(\textbf{s})=\sum_m c_{\textbf{k},\textbf{G}_m}e^{i(\textbf{G}_m+\textbf{k})\cdot\textbf{s}}
\\
\delta\psi^{\kappa\beta}_{n\textbf{k},\textbf{q}}(\textbf{s})=\sum_m \delta c_{\textbf{k}+\textbf{q},\textbf{G}_m}e^{i(\textbf{G}_m+\textbf{k}+\textbf{q})\cdot\textbf{s}},
\end{split}
\end{equation}
where we have dropped the band index and the $\kappa\beta$ indices
for the expansion coefficients $c$ and $\delta c$, and $m$ indexes a
reciprocal lattice vector $\textbf{G}_m$. Then, applying the momentum
operator,
\begin{equation}
\begin{split}
P^{\text{loc}}_{\alpha,\kappa\beta}(\textbf{G}+\textbf{q})
&=-\frac{2}{N_k}\sum_{\textbf{k}}\sum_{m,m^\prime}\int d^3r c^*_{\textbf{k},\textbf{G}_m}\delta c_{\textbf{k}+q_\alpha,\textbf{G}_{m^\prime}}\Big[(k_\alpha+q_\alpha+G_{\alpha,m^\prime}) e^{-i(\textbf{G}+\textbf{G}_m-\textbf{G}_{m^\prime})\cdot\textbf{r}}
\\
&\phantom{=-\frac{2}{N_k}\sum_{\textbf{k}}\sum_{m,m^\prime}\int d^3r c^*_{\textbf{k},\textbf{G}_m}\delta c_{\textbf{k}+q_\alpha,\textbf{G}_{m^\prime}}\Big[}+(k_\alpha+G_{\alpha,m}) e^{-i(\textbf{G}+\textbf{G}_m-\textbf{G}_{m^\prime})\cdot\textbf{r}}\Big]
\\
&=-\frac{4}{N_k}\sum_{\textbf{k}}\sum_{m} c^*_{\textbf{k},\textbf{G}_m}\delta c_{\textbf{k}+q_\alpha,\textbf{G}_{m}+\textbf{G}}\left(k_\alpha+G_{\alpha,m}+\frac{q_\alpha+G_\alpha}{2}\right)
\\
&=-\frac{2}{N_k}\sum_{n\textbf{k}}\langle u_{n\textbf{k}}\vert  e^{-i\textbf{G}\cdot\hat{\textbf{r}}}\left(\hat{p}^{\textbf{k}}_\alpha+\frac{q_\alpha}{2}\right) +\left(\hat{p}^{\textbf{k}}_\alpha+\frac{q_\alpha}{2}\right) e^{-i\textbf{G}\cdot\hat{\textbf{r}}}\vert\delta u^{\kappa\beta}_{n\textbf{k},\textbf{q}}\rangle,
\end{split}
\end{equation}
where, in the last line, we have restored the band and $\kappa\beta$
indices, $\hat{p}_\alpha^{\textbf{k}}=-i\hat{\nabla}_\alpha+k_\alpha$
is the cell-periodic momentum operator ($\hat{\nabla}_\alpha$ is a
spatial derivative in the $\alpha$ direction), and we have used that
$\psi_{n\textbf{k}}(\textbf{s})=u_{n\textbf{k}}(\textbf{s})e^{i\textbf{k}\cdot\textbf{s}}$.
In Sec.~\ref{Bench}, we use this result to calculate real-space
moments of the local contribution to the FxE coefficient. Otherwise,
we are usually interested in the $\textbf{G}=0$ term:
\begin{equation}
\begin{split}
\label{pkq}
\overline{P}^{\textbf{q},\text{loc}}_{\alpha,\kappa\beta}
&=-\frac{4}{N_k}\sum_{n\textbf{k}}\langle u_{n\textbf{k}}\vert\left(\hat{p}^{\textbf{k}}_\alpha+\frac{q_\alpha}{2}\right)\vert\delta u^{\kappa\beta}_{n\textbf{k},\textbf{q}}\rangle.
\end{split}
\end{equation}

\section{Current density in the presence of nonlocal pseudopotentials\label{JNL}}

Here we derive 
the contributions to the current from the nonlocal potentials
[$P^{\textbf{q},\text{nl}}_{\alpha,\kappa\beta}$ in
  Eq.~(\ref{PqExpand})], which we obtain by 
expanding the nonlocal current-density operator
up to second order in \textbf{q} [Eq.~(\ref{JqExpand})],
\begin{equation}
\label{Pqexpand2}
\begin{split}
\overline{P}_{\alpha,\kappa\beta}^{\textbf{q},\text{nl}}&\simeq \frac{4}{N_k}\sum_{n\textbf{k}}\Bigg[\langle u_{n\textbf{k}}\vert \hat{\mathcal{J}}_\alpha^{\textbf{k},\text{nl}(0)}  \vert\delta u^{\kappa\beta}_{n\textbf{k},\textbf{q}}\rangle
+\frac{1}{2}\sum_{\gamma=1}^3 q_\gamma\langle u_{n\textbf{k}}\vert \hat{\mathcal{J}}_{\alpha,\gamma}^{\textbf{k},\text{nl}(1)}  \vert\delta u^{\kappa\beta}_{n\textbf{k},\textbf{q}}\rangle
+\frac{1}{6}\sum_{\gamma=1}^3\sum_{\xi=1}^3q_\gamma q_\xi\langle u_{n\textbf{k}}\vert \hat{\mathcal{J}}_{\alpha,\gamma \xi}^{\textbf{k},\text{nl}(2)}  \vert\delta u^{\kappa\beta}_{n\textbf{k},\textbf{q}}\rangle\Bigg],
\end{split}
\end{equation}

The nonlocal potential that we are interested in is that of the
norm-conserving pseudopotential. In reciprocal space, the nonlocal
potential in the separable Kleinman-Bylander\cite{Kleinman1982} form
is given by\cite{Martin2004}
\begin{equation}
\label{VNL}
  V^{\text{nl}}(\textbf{K},\textbf{K}^\prime)=\sum_{\zeta}e^{-i\textbf{K}\cdot\textbf{R}_\zeta}\left(\sum_{lm}\frac{Y_{\zeta lm}^*(\hat{\textbf{K}})T_{\zeta l}^{*}(\vert \textbf{K}\vert)\times T_{\zeta l}(\vert \textbf{K}^\prime\vert)Y_{\zeta lm}(\hat{\textbf{K}}^\prime)}{E^{\text{KB}}_{\zeta l}}\right)e^{i\textbf{K}^\prime \cdot\textbf{R}_\zeta}
\end{equation}
where $\textbf{K}=\textbf{G}+\textbf{k}$; \textbf{R}$_\zeta$ is the
atomic position of atom $\zeta$; $Y_{\zeta lm}$ is the spherical
harmonic for the $lm$ angular momentum channel; $T_{\zeta l}(K)$ is
the Fourier transform of the radial function, $\widetilde{\psi}_{\zeta
  l}(r)V_{\zeta l}(r)$, where $V_{\zeta l}(r)$ are the
pseudopotentials and $\widetilde{\psi}_{\zeta l}(r)$ the
pseudoorbitals; $E^{\text{KB}}_{\zeta
  l}=\langle\widetilde{\psi}_{\zeta l}\vert
\hat{V}_l\vert\widetilde{\psi}_{\zeta l}\rangle$ is the
Kleinman-Bylander energies. The term in the parentheses is the
nonlocal form factor, and the phase factors surrounding it are the
structure factors. We define
\begin{equation}
\langle\textbf{K}\vert \phi_{\zeta lm}\rangle=e^{i\textbf{K}\cdot\textbf{R}_\zeta}Y_{\zeta lm}(\hat{\textbf{K}})T_{\zeta l}(\vert \textbf{K}\vert)
\end{equation}
so 
\begin{equation}
\label{NLop}
\hat{V}^{\text{nl}}=\sum_{\zeta lm}\frac{\vert \phi_{\zeta lm}\rangle\langle \phi_{\zeta lm}\vert}{E^{\text{KB}}_{\zeta l}} .
\end{equation}

\subsection{Ismail-Beigi, Chang, and Louie straight-line path\label{ICL}}
For the straight-line path of Essin \textit{et al.}\cite{Essin2010} and Ismail-Beigi, Chang, and Louie,\cite{ICL2001} we combine Eq.~(\ref{JkqICL}) and (assuming we have TRS) Eq.~(\ref{PqTR}). Since we have already addressed the local part in Appendix \ref{Jloc}, we only consider the nonlocal part of the Hamiltonian, defining the operator
\begin{equation}
\begin{split}
\langle\textbf{s}\vert\hat{\mathcal{J}}_{\alpha}^{\textbf{k},\textbf{q},\text{ICL,nl}}\vert\textbf{s}^\prime\rangle&=-iV^{\textbf{k},\text{nl}}(\textbf{s},\textbf{s}^\prime)(s_\alpha-s_\alpha^\prime)\left[\frac{e^{-i\textbf{q}\cdot(\textbf{s}-\textbf{s}^\prime)}-1}{i\textbf{q}\cdot(\textbf{s}-\textbf{s}^\prime)}\right].
\end{split}
\end{equation}
Expanding the term in square brackets in powers of \textbf{q} gives
\begin{equation}
\begin{split}
\langle\textbf{s}\vert\hat{\mathcal{J}}_{\alpha}^{\textbf{k},\textbf{q},\text{ICL,nl}}\vert\textbf{s}^\prime\rangle&=iV^{\textbf{k},\text{nl}}(\textbf{s},\textbf{s}^\prime)(s_\alpha-s_\alpha^\prime)\left[1-\frac{i\textbf{q}\cdot(\textbf{s}-\textbf{s}^\prime)}{2}+\frac{[i\textbf{q}\cdot(\textbf{s}-\textbf{s}^\prime)]^2}{6}-\cdots\right].
\\
&=i\left[\hat{r}_\alpha,\hat{V}^{\textbf{k},\text{nl}}\right]_{\textbf{s}\textbf{s}^\prime}
-\frac{1}{2}\sum_{\gamma=1}^3 q_\gamma\left[\hat{r}_\gamma,\left[\hat{r}_\alpha,\hat{V}^{\textbf{k},\text{nl}}\right]\right]_{\textbf{s}\textbf{s}^\prime}
-\frac{i}{6}\sum_{\gamma=1}^3\sum_{\xi=1}^3 q_\gamma q_\xi\left[\hat{r}_\xi,\left[\hat{r}_\gamma,\left[\hat{r}_\alpha,\hat{V}^{\textbf{k},\text{nl}}\right]\right]\right]_{\textbf{s}\textbf{s}^\prime}+\cdots,
\end{split}
\end{equation}
so we can write the operator as
\begin{equation}
\begin{split}
\label{kgradop}
\hat{\mathcal{J}}_{\alpha}^{\textbf{k},\textbf{q},\text{ICL,nl}}&=\sum_{\gamma_1\cdots \gamma_n}\frac{q_{\gamma_1}\cdots q_{\gamma_n}}{(n+1)!} \hat{\mathcal{J}}_{\alpha,\gamma_1\cdots\gamma_n}^{\textbf{k},\text{ICL,nl}(n)}, \qquad
 \hat{\mathcal{J}}_{\alpha,\gamma_1\cdots\gamma_n}^{\textbf{k},\text{ICL,nl}(n)} = -\frac{\partial^{n+1}\hat{V}^{\textbf{k},{\text{nl}}}}{\partial k_\alpha \partial k_{\gamma_1}\cdots \partial k_{\gamma_n}}.
\end{split}
\end{equation}
In terms of the cell-periodic projectors $\phi_{\zeta
  lm}^\textbf{k}(\textbf{s})=e^{-i\textbf{k}\cdot\textbf{s}}\phi_{\zeta
  lm}(\textbf{s})$ [see Eq.~(\ref{NLop})], the lowest-order
  terms in Eq.~(\ref{kgradop}), to be incorporated into
  Eq.~(\ref{Pqexpand2}), are
\begin{equation}
\label{ICL0}
\hat{\mathcal{J}}_\alpha^{\textbf{k},\text{nl}(0)} =-\sum_{\zeta lm}\frac{1}{E^{\text{KB}}_{\zeta l}}\left(\vert \phi^{\textbf{k}}_{\zeta lm}\rangle\langle\partial_\alpha \phi^{\textbf{k}}_{\zeta lm}\vert+\vert\partial_\alpha \phi^{\textbf{k}}_{\zeta lm}\rangle\langle \phi^{\textbf{k}}_{\zeta lm}\vert\right),
\end{equation}
\begin{equation}
\label{ICL1}
\begin{split}
\hat{\mathcal{J}}_{\alpha,\gamma}^{\textbf{k},\text{ICL,nl}(1)}
=-\sum_{\zeta lm}\frac{1}{E^{\text{KB}}_{\zeta l}}\left(\vert\partial_\gamma \phi^{\textbf{k}}_{\zeta lm}\rangle\langle\partial_\alpha \phi^{\textbf{k}}_{\zeta lm}\vert
+\vert \phi^{\textbf{k}}_{\zeta lm}\rangle\langle\partial_\alpha\partial_\gamma \phi^{\textbf{k}}_{\zeta lm}\vert
+\vert\partial_\alpha\partial_\gamma \phi^{\textbf{k}}_{\zeta lm}\rangle\langle \phi^{\textbf{k}}_{\zeta lm}\vert
+\vert \partial_\alpha \phi^{\textbf{k}}_{\zeta lm}\rangle\langle\partial_\gamma \phi^{\textbf{k}}_{\zeta lm}\vert\right),
\end{split}
\end{equation}
and
\begin{equation}
\label{ICL2}
\begin{split}
\hat{\mathcal{J}}_{\alpha,\gamma\xi}^{\textbf{k},\text{ICL,nl}(2)}
=-\sum_{\zeta lm}\frac{1}{E^{\text{KB}}_{\zeta l}}&\big(\vert\partial_\xi\partial_\gamma \phi^{\textbf{k}}_{\zeta lm}\rangle\langle\partial_\alpha \phi^{\textbf{k}}_{\zeta lm}\vert
+\vert\partial_\xi \phi^{\textbf{k}}_{\zeta lm}\rangle\langle\partial_\gamma\partial_\alpha \phi^{\textbf{k}}_{\zeta lm}\vert
+\vert \partial_\gamma \phi^{\textbf{k}}_{\zeta lm}\rangle\langle\partial_\alpha\partial_\xi \phi^{\textbf{k}}_{\zeta lm}\vert
+\langle  u_{n\textbf{k}}\vert \phi^{\textbf{k}}_{\zeta lm}\rangle\langle\partial_\gamma\partial_\alpha\partial_\xi \phi^{\textbf{k}}_{\zeta lm}\vert
\\
&+\vert\partial_\gamma\partial_\alpha\partial_\xi \phi^{\textbf{k}}_{\zeta lm}\rangle\langle \phi^{\textbf{k}}_{\zeta lm}\vert
+\langle  u_{n\textbf{k}}\vert\partial_\alpha\partial_\xi \phi^{\textbf{k}}_{\zeta lm}\rangle\langle\partial_\gamma \phi^{\textbf{k}}_{\zeta lm}\vert
+\vert \partial_\gamma\partial_\alpha \phi^{\textbf{k}}_{\zeta lm}\rangle\langle\partial_\xi \phi^{\textbf{k}}_{\zeta lm}\vert
+\vert \partial_\alpha \phi^{\textbf{k}}_{\zeta lm}\rangle\langle\partial_\gamma\partial_\xi \phi^{\textbf{k}}_{\zeta lm}\vert\big).
\end{split}
\end{equation}
These correspond to last three terms in Eq.~(\ref{JqExpand}), here
specialized to the ICL path.  Note that
$\hat{\mathcal{J}}_\alpha^{\textbf{k},\text{nl}(0)} = -{\partial
  \hat{V}^{\textbf{k},\text{nl}}}/{\partial k_\alpha}$ represents the
well-known nonlocal correction to the Born effective charge (with an
overall negative sign from the electron charge), which, combined with
the local part [Eq.~(\ref{pkq})] yields the velocity operator
$\hat{v}_\alpha^{\textbf{k},\textbf{q}}$ and should be unsensitive to the path choice.

\subsection{Pickard and Mauri path through atom center\label{PM}}

The PM \cite{Pickard2003} path goes through the center of the
atom. For simplicity of the derivation, we consider a single atom
positioned at the origin ($\textbf{R}=0$); the generalization to an
atom not at the origin simply involves an extra phase in the structure
factors in Eq.~(\ref{VNL}). Then Eq.~(\ref{JqPMNL}) becomes
\begin{equation}
\begin{split}
\langle\textbf{s}\vert\hat{\mathcal{J}}_{\alpha}^{\text{PM,nl}}(\textbf{q})\vert\textbf{s}^\prime\rangle=-iV^{\text{nl}}(\textbf{s},\textbf{s}^\prime)&\bigg[s^\prime_\alpha\frac{1-e^{-i\textbf{q}\cdot\textbf{s}^\prime}}{i\textbf{q}\cdot\textbf{s}^\prime}+s_\alpha\frac{e^{-i\textbf{q}\cdot\textbf{s}}-1}{i\textbf{q}\cdot\textbf{s}}\bigg].
\end{split}
\end{equation}
Following the same steps as in Appendix \ref{ICL}, we arrive at slightly
different current operators for the terms to first and second
order in \textbf{q} (the zeroth order term is the same as for the ICL
path, as expected),
\begin{equation}
\begin{split}
\label{PM1}
\hat{\mathcal{J}}^{\textbf{k},\text{PM,nl}(1)}_{\alpha,\gamma}=-\sum_{\zeta lm}\frac{1}{E^{\text{KB}}_{\zeta l}}\big(2\vert\partial_\alpha\phi^{\textbf{k}}_{\zeta lm}\rangle\langle\partial_\gamma\phi^{\textbf{k}}_{\zeta lm}\vert
+\vert\phi^{\textbf{k}}_{\zeta lm}\rangle\langle\partial_\alpha\partial_\gamma\phi^{\textbf{k}}_{\zeta lm}\vert
+\vert\partial_\alpha\partial_\gamma\phi^{\textbf{k}}_{\zeta lm}\rangle\langle\phi^{\textbf{k}}_{\zeta lm}\vert\big),
\end{split}
\end{equation}
\begin{equation}
\begin{split}
\label{PM2}
\hat{\mathcal{J}}^{\textbf{k},\text{PM,nl}(2)}_{\alpha,\gamma\xi}=-\sum_{\zeta lm}\frac{1}{E^{\text{KB}}_{\zeta l}}\big(3\vert\partial_\alpha\partial_\gamma\phi^{\textbf{k}}_{\zeta lm}\rangle\langle\partial_\xi\phi^{\textbf{k}}_{\zeta lm}\vert
+3\vert\partial_\alpha\phi^{\textbf{k}}_{\zeta lm}\rangle\langle\partial_\gamma\partial_\xi\phi^{\textbf{k}}_{\zeta lm}\vert
+\vert\phi^{\textbf{k}}_{\zeta lm}\rangle\langle\partial_\alpha\partial_\gamma\partial_\xi\phi^{\textbf{k}}_{\zeta lm}\vert
+\vert\partial_\alpha\partial_\gamma\partial_\xi\phi^{\textbf{k}}_{\zeta lm}\rangle\langle\phi^{\textbf{k}}_{\zeta lm}\vert\big).
\end{split}
\end{equation}
We see immediately that, for the case of a longitudinal perturbation,
 Eqs.~(\ref{PM1}) and (\ref{PM2}) are
identical to their ICL counterparts [\textit{cf.} Eq.~(\ref{ICL1}) and
Eq.~(\ref{ICL2})].

\section{Diamagnetic correction\label{Appdiamag}}
In this section we provide some details about the calculation of the
CRG contribution to the transverse and shear FxE coefficients,
which is related to the diamagnetic susceptibility. We refer the
reader to Ref.~\onlinecite{StengelUNPUB} for a complete discussion.

For the case of a small deformation \textbf{u} that is applied to the
atoms of a crystal adiabatically through the perturbation parameter
$\lambda(t)$, the CRG contribution to linear order in the velocity
is
\begin{equation}
\label{hdyn}
\dot{\lambda}\hat{H}^{(\dot{\lambda})}=-\frac{1}{2}\left(\hat{\textbf{A}}\cdot\hat{\textbf{p}}+\hat{\textbf{p}}\cdot\hat{\textbf{A}}\right).
\end{equation}
Here $\bf A$ is not the vector potential of electromagnetism, but one
that emerges when transforming from the static reference frame to the
CRG one.  For a monochromatic perturbation, it becomes just
$\textbf{A}=\dot{\lambda}\textbf{u}=\dot{\lambda}e^{i\textbf{q}\cdot\textbf{r}}$,
so
\begin{equation}
\begin{split}
\label{hdyn2}
\hat{H}^{(\dot{\lambda}_\beta)}(\textbf{q})&=-e^{i\textbf{q}\cdot\hat{\textbf{r}}}\left(\hat{p}_\beta+\frac{q_\beta}{2}\right)
\end{split}
\end{equation}
which we recognize as the local current operator [\textit{cf.}
Eq.~(\ref{LocNL4}) or (\ref{pkq})]. Therefore, the first-order,
cell-periodic wavefunctions with respect to this perturbation are
\begin{equation}
\label{udyn}
\vert\partial_{\dot{\lambda}_\beta}u^\beta_{n\textbf{k},\textbf{q}}\rangle=\sum_m^{\text{unocc}}\frac{\vert u_{m\textbf{k},\textbf{q}}\rangle\langle u_{m\textbf{k},\textbf{q}}\vert\left(\hat{p}_\beta^{\textbf{k}}+q_\beta/2\right) \vert  u_{n\textbf{k}}\rangle}{\epsilon_{m\textbf{k},\textbf{q}}-\epsilon_{n\textbf{k}}},
\end{equation}
and the (cell averaged) induced polarization from the CRG part of
the metric perturbation is
\begin{equation}
\label{pdyn}
\begin{split}
\overline{P}^{\textbf{q},\text{ CRG}}_{\alpha,\beta}&=\frac{4}{N_k}\sum_{n\textbf{k}} \langle u_{n\textbf{k}}\vert\hat{\mathcal{J}}^{\textbf{k},\textbf{q}}_\alpha\vert\partial_{\dot{\lambda}_\beta}u^\beta_{n\textbf{k},\textbf{q}}\rangle.
\end{split}
\end{equation}
The contribution to the FxE coefficient is determined by taking the
second derivative of
$\overline{P}^{\textbf{q},\text{CRG}}_{\alpha,\beta}$ with respect to
$q$:
\begin{equation}
\label{pdyn2}
\begin{split}
\overline{P}^{(2,\omega\nu),\text{ CRG}}_{\alpha,\beta}&=\frac{\partial^2\overline{P}^{\textbf{q},\text{CRG}}_{\alpha,\beta}}{\partial q_\omega \partial q_\nu}\Bigg\vert_{\textbf{q}=0}.
\end{split}
\end{equation}
The CRG contribution is closely related to the diamagnetic
susceptibility, $\chi_{\alpha\beta}$. In fact, in the case where only
local potentials are present in the Hamiltonian [so that
$\hat{\mathcal{J}}^{\textbf{k},\textbf{q}}_\beta=-(\hat{p}_\beta^{\textbf{k}}+q_\beta/2)$
in Eq.~(\ref{pdyn})], Eq.~(\ref{pdyn2}) has the same form as the
expressions for the magnetic susceptibility derived in, e.g.,
Refs.~\onlinecite{Vignale1991} and ~\onlinecite{Mauri1996}
[\textit{cf.} Eq.~(11) and Eq.~(9) in those works, respectively].

The magnetic susceptibility relates the magnetization, ${\bf M}$, to the external magnetic field, ${\bf B}$, 
via $M_\gamma=\chi_{\gamma\lambda}^{\text{mag}}B_\lambda$. This can be rewritten to relate the bound currents to the vector potential,
\begin{equation}
J_\alpha=\epsilon^{\alpha\zeta\gamma}\nabla_\zeta\chi_{\gamma\lambda}\epsilon^{\lambda\rho\beta}\nabla_\rho A_\beta,
\end{equation}
so that
\begin{equation}
\overline{P}^{\textbf{q},\text{ CRG}}_{\alpha,\beta}\sim\epsilon^{\alpha\zeta\gamma}q_\zeta\chi_{\gamma\lambda}\epsilon^{\beta\lambda\rho}q_\rho,
\end{equation}
where we have expressed the spatial derivatives in reciprocal space
and canceled the resulting negative sign by permutating the second
Levi-Civita symbol. Performing the ${\bf q}$-derivatives in
Eq.~(\ref{pdyn2}) gives
\begin{equation}
\label{apchi}
\begin{split}
\overline{P}^{(2,\omega\nu),\text{ CRG}}_{\alpha,\beta}&=\sum_{\gamma\lambda}\left(\epsilon^{\alpha\omega\gamma}\epsilon^{\beta\lambda\nu}+\epsilon^{\alpha\nu\gamma}\epsilon^{\beta\lambda\omega}\right)\chi_{\gamma\lambda}.
\end{split}
\end{equation}

In the case that nonlocal potentials are present in the Hamiltonian, a
calculation of the magnetic susceptibility would involve replacing the
``displacement velocity'' operator,
$-\left(\hat{p}_\beta^{\textbf{k}}+q_\beta/2\right)$, in
Eq.~(\ref{udyn}) with the full electromagnetic current
operator from Eq.~(\ref{JqExpand}), as well as evaluating extra terms
originating from the second-order Hamiltonian
\cite{ICL2001,Pickard2001,Pickard2003}.
This is in contrast to the case of the CRG contribution we would
like to calculate, where the only change in the case of nonlocal
potentials is replacing
$\hat{\mathcal{J}}^{\textbf{k},\textbf{q}}_\alpha$ in Eq.~(\ref{pdyn})
with the full current operator from Eq.~(\ref{JqExpand});
Eqs.~(\ref{hdyn2}) and (\ref{udyn}) are unchanged. Therefore,
Eq.~(\ref{pdyn}) does not strictly correspond to the magnetic
susceptibility in this case. However, we show in Sec.~\ref{Disc} that
the numerical values are quite similar to previously calculated
diamagnetic susceptibilities .

\section{Divergence of the current at the atomic site for the PM path\label{appDiv}}
To illustrate the point that nonlocal pseudopotentials allow
unphysical transfer of charge between \textbf{r} and
$\textbf{r}^\prime$, we shall consider the PM\cite{Pickard2003}
definition of the current density, which provides a particularly
transparent manifestation of such unphysical behavior.
For simplicity, we focus our attention on a single atomic sphere [so
we drop the $\zeta$ index of Eq.~(\ref{HPM})], and we set the
corresponding nuclear site as the coordinate origin. (There is no
approximation here, as the contributions from different sites are
spatially separated and additive.)
Now suppose we wish to evaluate the nonlocal current density at the
point ${\bf r}_0$. We need then to calculate Eq.~(\ref{dHdAq}) with
Eq.~(\ref{HPM}), using a Dirac delta as a vector potential,
\begin{equation}
{\bf A}({\bf r}) = A \hat{\bf r}_0 \, \delta ({\bf r - r}_0) = A \hat{\bf r}_0\delta(\hat{\textbf{r}}-\hat{\textbf{r}}_0)\frac{\delta(r-r_0)}{4\pi r^2},
\end{equation}
where the caret above the position variable denotes a direction (not
to be confused with the position operator), and in the second equality we have written the Dirac delta function in spherical coordinates. We choose the vector potential to be
oriented along the radial direction, as this is the only allowed
component within the PM theory: it is easy to see that a
purely tangential ${\bf A}$ field yields a vanishing nonlocal
contribution to the current [see Eq.~(\ref{HPM})].
Then, the line integral needed for the first order term in Eq.~(\ref{HA2}) is
\begin{equation}
\begin{split}
\int_{\textbf{s}^\prime\rightarrow 0 \rightarrow\textbf{s}}\textbf{A}\cdot d\ell &=\int_0^1\textbf{A}(\tau\textbf{s})\cdot\textbf{s}d\tau-\int_0^1 \textbf{A}(\tau\textbf{s}^\prime)\cdot\textbf{s}^\prime d\tau
\\
 &=A\hat{\textbf{r}}_0\cdot\left[\delta(\hat{\textbf{s}}-\hat{\textbf{r}}_0)\textbf{s}\int^1_0 \frac{\delta(\tau s-r_0)}{4\pi (s\tau)^2}d\tau-\delta(\hat{\textbf{s}}^\prime-\hat{\textbf{r}}_0)\textbf{s}^\prime\int^1_0 \frac{\delta(\tau s^\prime-r_0)}{4\pi (s^\prime\tau)^2}d\tau\right]
\\
 &=\frac{A}{4\pi r_0^2}\left[\delta(\hat{\bf s} - \hat{\bf r}_0)\theta(s-r_0)-\delta(\hat{\bf s}^\prime - \hat{\bf r}_0)\theta(s^\prime-r_0)\right],
\end{split}
\end{equation}
where $\theta$ is the Heaviside step function. Therefore, we can write
the current-density operator as (recall that the tangential components
vanish, so the current is purely radial)
\begin{equation}
\label{radJ}
\langle\textbf{s}\vert\hat{\mathcal{J}}(\textbf{r})\vert\textbf{s}^\prime\rangle=\frac{iV^{\text{nl}}(\textbf{s},\textbf{s}^\prime)}{4\pi r^2}\left[\delta(\hat{\bf s} - \hat{\bf r})\theta(s-r)-\delta(\hat{\bf s}^\prime - \hat{\bf r})\theta(s^\prime-r)\right] .
\end{equation}
Considering a general time-dependent wavefunction as in
Eq.~(\ref{Js}), the current density is
\begin{equation}
\label{JdivPM}
\begin{split}
\textbf{J}^{\text{nl}}(\textbf{r},t)&=\frac{i\hat{\textbf{r}}}{4\pi r^2}\int d^3s \int d^3s^\prime \Psi^*(\textbf{s},t)\Psi(\textbf{s}^\prime,t)V^{\text{nl}}(\textbf{s},\textbf{s}^\prime)\left[\delta(\hat{\bf s} - \hat{\bf r})\theta(s-r)-\delta(\hat{\bf s}^\prime - \hat{\bf r})\theta(s^\prime-r)\right]
\\
&=\frac{i\hat{\textbf{r}}}{r^2}\sum_{lm}\int_r^\infty ds\left[\langle\phi_{lm}\vert\Psi(t)\rangle \Psi^*(s\hat{\textbf{r}},t)\phi_{lm}(s\hat{\textbf{r}})-\langle\Psi(t)\vert\phi_{lm}\rangle\Psi(s\hat{\textbf{r}},t)\phi_{lm}^*(s\hat{\textbf{r}})\right]s^2
\\
&=\frac{\hat{\textbf{r}}}{r^2}\int_r^\infty ds \,\rho^{\text{nl}}(s\hat{\textbf{r}})s^2
\end{split}
\end{equation} 
where we have identified the nonlocal charge
$\rho^{\text{nl}}(\textbf{r})=-i\langle\Psi\vert\left[\vert
  \textbf{r}\rangle\langle
  \textbf{r}\vert,\hat{V}^{\text{nl}}\right]\vert\Psi\rangle$
[\textit{cf.} Eq.~(\ref{rhoNL})].
Note that the upper limit of the integral can be set to $r_{\rm c}$,
i.e., the core radius that was used in the generation of the
pseudopotential (the nonlocal current density field is strictly
contained within a sphere of radius $r_{\rm c}$). This shows that, in
the special case of the Pickard-Mauri theory, the nonlocal density
does, in fact, provide complete information about the current density.

Unfortunately, a consequence of the above derivations is that the
nonlocal current density \emph{diverges} as $|{\bf r - R}|^{-2}$ in
the vicinity of an atomic site ${\bf R}$.
To see this it suffices to observe that the integral in the above equation
tends, for $r \rightarrow 0$, to a direction-dependent constant,
\begin{equation}
\label{Cr}
\int_0^{+\infty} s^2 ds \, \rho^{\rm nl}(s \hat{\bf r}) = C(\hat{\bf r}).
\end{equation}
Thus, the current-density field diverges near the atomic site as
\begin{equation}
{\bf J}^{\rm nl}({\bf r}) \sim \frac{\hat{\bf r} C(\hat{\bf r})}{r^2} .
\end{equation}

\end{widetext}

\bibliography{flexo}
\end{document}